\documentclass[twocolumn,prl,superscriptaddress]{revtex4}
\usepackage{amsmath}
\usepackage{graphicx}
\usepackage{amssymb}
\usepackage{dcolumn}
\usepackage{mathrsfs}
\usepackage{bm}
\makeatletter

\newcommand{\Rmnum}[1]{\expandafter\@slowromancap\romannumeral #1@}
\makeatother

\begin{document}

\title{Maxwell Quasiparticles Emerged in Optical Lattices}
\author{Yan-Qing Zhu}
\affiliation{National Laboratory of Solid State Microstructures
and School of Physics, Nanjing University, Nanjing 210093, China}

\author{Dan-Wei Zhang}
\email{zdanwei@126.com}\affiliation{Guangdong Provincial Key Laboratory of Quantum Engineering and Quantum Materials,
SPTE, South China Normal University, Guangzhou 510006, China}

\author{Hui Yan}
\affiliation{Guangdong Provincial Key Laboratory of Quantum
Engineering and Quantum Materials, SPTE, South China Normal
University, Guangzhou 510006, China}

\author{Ding-Yu Xing}
\affiliation{National Laboratory of Solid State Microstructures
and School of Physics, Nanjing University, Nanjing 210093, China}
\affiliation{ Collaborative Innovation Center of Advanced
Microstructures, Nanjing 210093, China}

\author{Shi-Liang Zhu}
\email{slzhu@nju.edu.cn}
\affiliation{National Laboratory of Solid
State Microstructures and School of Physics, Nanjing University,
Nanjing 210093, China}
\affiliation{Synergetic Innovation Center
of Quantum Information and Quantum Physics, University of Science
and Technology of China, Hefei, Anhui 230026, China}

\date{\today}

\begin{abstract}
We construct a two-dimensional tight-binding model of an optical
lattice, where the low energy excitations should be described by
the spin-1 Maxwell equations in the Hamiltonian form, and such
linear dispersion excitations with pesudospin-1 are so called as
the Maxwell quasiparticles. The system has rich topological
features, for examples, the threefold degeneracy points called
Maxwell points may have nontrivial $\pm 2\pi$ Berry phases and the
anomalous quantum Hall effect with spin-momentum locking may appear in
topological Maxwell insulators. We propose realistic schemes for
realizing the Maxwell metals/insulators and detecting the
intrinsic properties of the topological Maxwell quasiparticles
with ultracold atoms in optical lattices.

\end{abstract}
\maketitle


\emph{Introduction.---} Discovery of new particles in nature or
new quasiparticles in condensed matter systems lies at the heart of
the modern physics \cite{Wilczek}. One of the recnet examples is
that the relativistic Dirac fermions emerged in a graphene have
attracted interest in condensed matter physics as well as in
quantum field theory \cite{Castro}. Furthermore, it was
demonstrated that Weyl fermions, the massless spin-1/2 particles
in quantum field theory but have never been observed as
fundamental particles in nature, can emerge as quasiparticles in
solids \cite{Wan,Xu,Lv} and photonic crystals \cite{Lu}.
Most interestingly, the Dirac and Weyl fermions have rich
topological features \cite{Hasan,Qi,Wan,Xu,Lv,Lu}. However, the
quasiparticles with higher spin numbers that are also fundamentally
important are rarely studied \cite{Bradlyn,Lan}. For instance, the
massless photons with spin-1 are fundamental particles in nature,
which are described by Maxwell equations. Note that the Dirac and Weyl
fermions have already been well investigated in the field of the cold
atoms \cite{Zhu,Tarruell,Dubcek,Zhang2012,Zhang2015,Duca}. A
natural question is raised: can we realize the Maxwell
quasiparticles (relativistic excitations with pesudospin-1) with a
well-designed ultracold atomic system?


In this Letter, we propose schemes to create and explore
Maxwell quasiparticles in a two-dimensional (2D) optical
lattice (OL). We first rewrite the Maxwell equations in an anisotropic medium
in the form of the Schr\"{o}dinger equation and then construct 2D
optical lattices, where the low energy excitations should be
described by the Maxwell Hamiltonian. By tuning the on-site
spin-flip parameter, we show that the system can have rich quantum phases:
topological or normal Maxwell insulator, topological or normal
Maxwell metal. The topological Maxwell metal is characterized with
the threefold degeneracy points, the so-called Maxwell points, which
have nontrivial $\pm 2\pi$ Berry phases. The low-energy
excitations near the Maxwell point behave like photons
described by the Maxwell equations in the form of the
Schr\"{o}dinger equation. Furthermore, we find nontrivial edge states with
spin-momentum locking in the topological Maxwell insulating phases, mimicking the circularly-polarized photons.
Our work reveals the topological properties of Maxwell quasiparticles, which are analogy with the Dirac and Weyl fermions
in topological insulators and topological semimetals.

\emph{Maxwell equations in the form of the Schr\"{o}dinger
equation.---} In a region in the absence of free charges and currents, the
well-known Maxwell equations in matter are given by
\begin{equation}
\begin{split}
\nabla\times\mathbf{E}&=-\frac{\partial{\mathbf{B}}}{\partial{t}},\quad
\nabla\cdot\mathbf{{E}}=0,\\
\nabla\times\mathbf{H}&=\
\frac{\partial{\mathbf{D}}}{\partial{t}},\quad\
\nabla\cdot\mathbf{B}=0,
\end{split}
\end{equation}
where the displacement field
$\mathbf{D}=\varepsilon_0\varepsilon_r\mathbf{E}$ with
$\mathbf{E}$ being the electric field, the magnetic field $\mathbf{B}=\mu_0\mu_r\mathbf{H}$ with $\mathbf{H}$ being the magnetizing field,
 $\varepsilon_r$ and $\mu_r$ are
the relative permittivity and permeability, respectively. In an
anisotropic medium, $\varepsilon_r$ and $\mu_r$ are tensors rather
than numbers. To simplify the proceeding analysis, we assume that
the tensors $\varepsilon_r$ and $\mu_r$ are simultaneously
diagonalized. We can define the photon wave function as
$\mathbf{\Phi}(\mathbf{r},t)=\mathbf{\tilde{E}}(\mathbf{r},t)+i\mathbf{\tilde{H}}(\mathbf{r},t)$
\cite{Oppenheimer,Good}, where
$\tilde{E}_\alpha=\sqrt{\varepsilon_0\varepsilon_\alpha}E_\alpha$
and $\tilde{H}_\alpha=\sqrt{\mu_0\mu_\alpha}H_\alpha$. Then the
Maxwell equations can be rewritten as $\nabla\cdot\mathbf{\Phi}=0$
and
$i\hbar\frac{\partial{\Phi^\alpha}}{\partial{t}}=\nu_{\alpha\gamma}(i\epsilon_{\alpha\beta\gamma})\hat{P}_\beta\frac{\partial{\Phi^\gamma}}{\partial\beta},$
where $\nu_{\alpha\gamma}=c/\sqrt{\varepsilon_\alpha\mu_\gamma}$,
$\hat{P}_\beta=-i\hbar\partial_{\beta}$, and
$\epsilon_{\alpha\beta\gamma}$ $(\alpha,\beta,\gamma=x,y,z)$ is
the Levi-Civita symbol. We can rewrite the Maxwell
equations in the form of the Schr\"{o}dinger's equation as
$i\hbar\frac{\partial}{\partial{t}}\mathbf{\Phi}=\hat{H}_M\mathbf{\Phi}$
where the Maxwell Hamiltonian of the photons is given by \cite{Supplement}
\begin{equation}
\label{MaxwellEq}
\hat{H}_M=v_x\hat{S}_x\hat{P}_x+v_y\hat{S}_y\hat{P}_y+v_z\hat{S}_z\hat{P}_z.
\end{equation}
Here $\hat{\mathbf{S}}=(\hat{S}_x,\hat{S}_y,\hat{S}_z)$ are the
spin matrices for a particle of spin-1 \cite{Supplement}. Equation
(\ref{MaxwellEq}) is analogous to the Dirac (Weyl) equation for the
massless relativistic fermions with spin-1/2. In this paper, we
demonstrate that the low-energy physics in some well-designed
OLs loaded with free fermions (bosons) should be
described by the Schr\"{o}dinger equation with the Maxwell
Hamiltonian (\ref{MaxwellEq}), and thus we call such
quasiparticles as the Maxwell fermions (bosons).

\begin{figure}[htbp]\centering
\includegraphics[width=8cm]{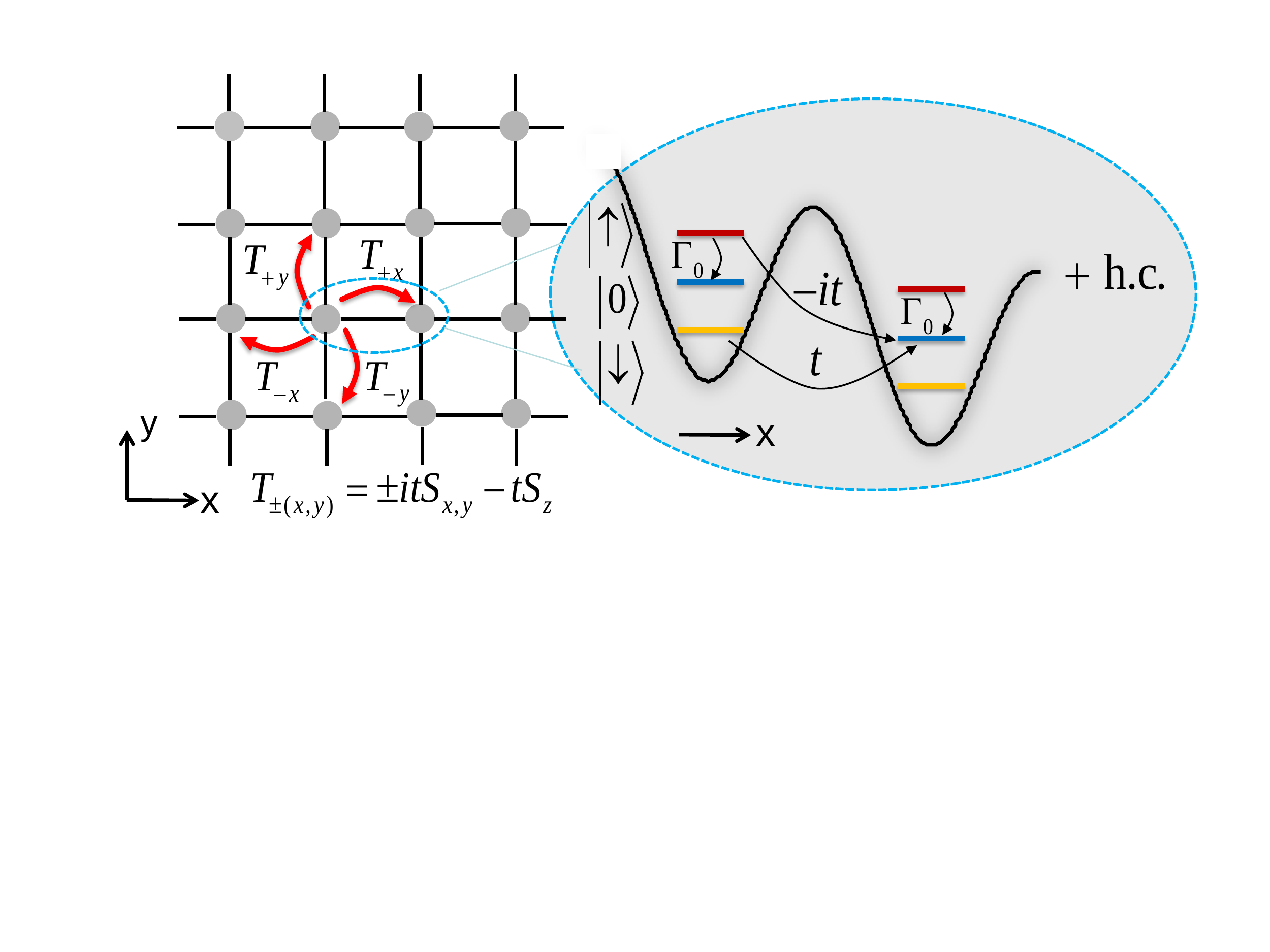}
 \caption{(color online). Schematic diagram of realizing the 2D Maxwell lattices. The model Hamiltonian (\ref{2DHam})
 can be realized with ultra-cold atoms loaded in a square OL with the spin-flip
 hopping $T_{\pm (x,y)}$ and on-site spin-flip term  $\Gamma_0$. The three atomic internal states $|\uparrow\rangle,|0\rangle,|\downarrow\rangle$
 form the (pseudo)spin-1 basis and the required atomic
 spin-flip hopping, such as $T_{+x}$ shown on the right, can be engineered by Raman lasers.} \label{OptLat}
\end{figure}

\emph{The model.---} The Maxwell quasiparticles can be realized
with two different schemes. We can use non-interacting fermionic
or bosonic atoms in a square OL and choose three
atomic internal states in the ground state manifold to encode the
three spin states $|s\rangle$ ($s=\uparrow,0,\downarrow$).
Alternatively, it can be realized by using single-component atoms
in OLs with three sublattices, where the pseudospin-1
basis are replaced by the three sublattices $A,B,C$ in a unit
cell. The detailed schemes are addressed in Supplemental Materials
\cite{Supplement}. For the conceptual simplicity, we discuss the
formal one in the main text. The model Hamiltonian we considered is
given by
\begin{equation} \label{2DHam}
\hat{H}=t\sum_{\mathbf{r}}\left[
\hat{H}_{\mathbf{rx}}+\hat{H}_{\mathbf{ry}}+  \left(
\Gamma_0\hat{a}^{\dag}_{\mathbf{r},0}\hat{a}_{\mathbf{r},\uparrow}
+\textrm{H.c.}\right) \right],
\end{equation}
where
$\hat{H}_{\mathbf{rx}}=-\hat{a}^{\dag}_{\mathbf{r-x},0}(\hat{a}_{\mathbf{r},
\downarrow}+i\hat{a}_{\mathbf{r},\uparrow}) +
\hat{a}^{\dag}_{\mathbf{r+x},0}(\hat{a}_{\mathbf{r},\downarrow}-i\hat{a}_{\mathbf{r},\uparrow})+\textrm{H.c.}$
and
$\hat{H}_{\mathbf{ry}}=\hat{a}^{\dag}_{\mathbf{r-y},\uparrow}(\hat{a}_{\mathbf{r},\downarrow}+i\hat{a}_{\mathbf{r},0})
-
\hat{a}^{\dag}_{\mathbf{r+y},\uparrow}(\hat{a}_{\mathbf{r},\downarrow}-i\hat{a}_{\mathbf{r},0})+\textrm{H.c.}$
respectively represent the spin-flip hopping along the $x$ and $y$
axis with the tunneling amplitude $t$, $\hat{a}_{\mathbf{r},s}$ is
the annihilation operator on site $\mathbf{r}$ for the spin state
$|s\rangle$, and $\Gamma_0=2iM$ with the tunable parameter $M$
being the strength of the on-site spin-flip.  The required
spin-flip hopping and on-site spin-flip terms are demonstrated in
Fig. \ref{OptLat}. The spin-flip hopping terms
$\hat{H}_{\mathbf{rx}}$ and $\hat{H}_{\mathbf{ry}}$ can be
achieved by the Raman-assisted tunneling scheme
\cite{LAT1,LAT2,LAT3,Dalibard,Goldman} with Raman lasers, which
address atoms with the laser-frequency and polarization selections
\cite{Supplement}. The on-site spin-flip term
$\Gamma_0\hat{a}^{\dag}_{\mathbf{r},0}\hat{a}_{\mathbf{r},\uparrow}$
can be achieved and tuned by application of a simple
radio-frequency field or additional Raman beams.

\begin{figure}[htbp]\centering
\includegraphics[width=8.5cm]{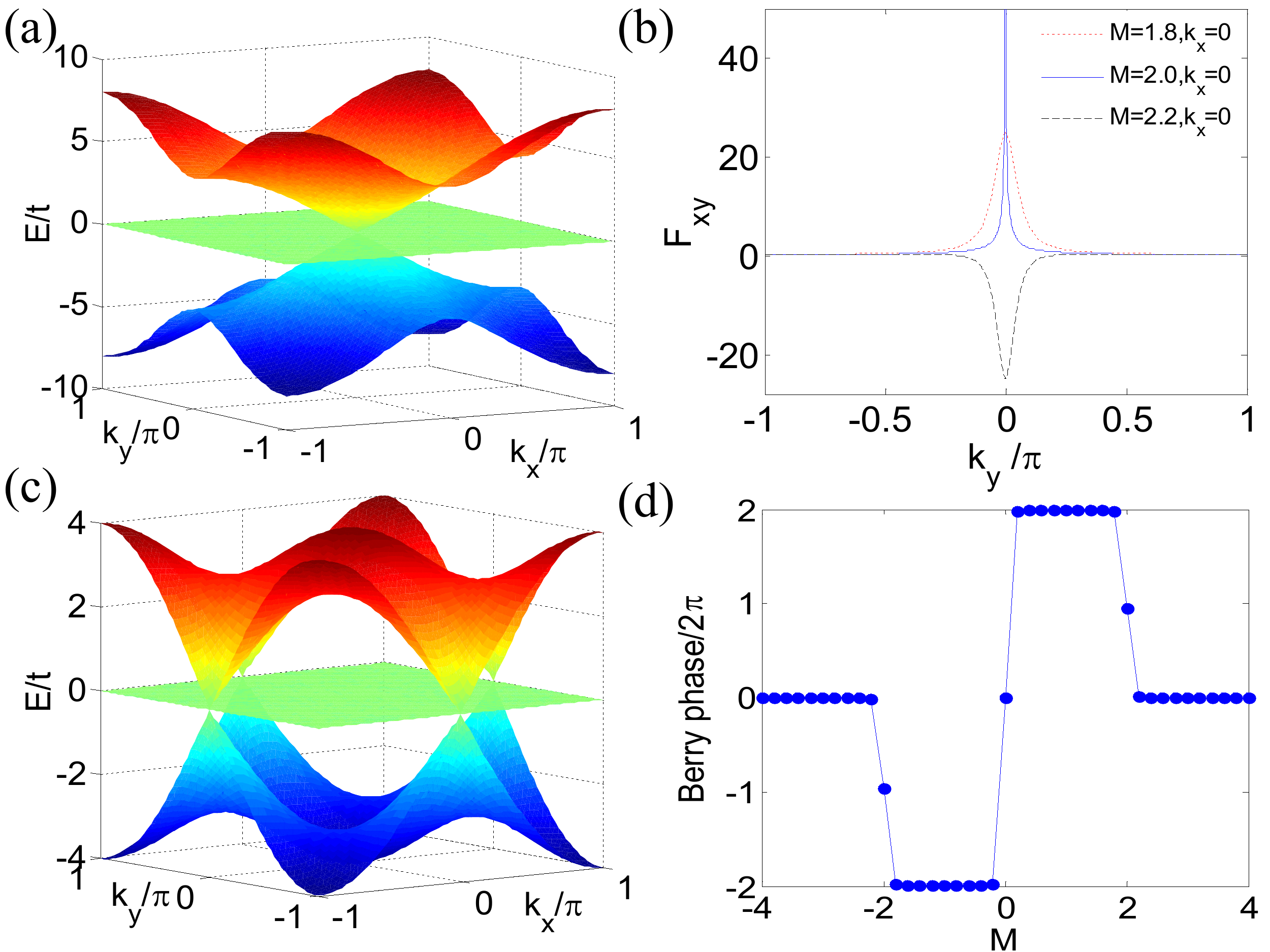}
\caption{ (color online). The energy spectra and topological
features of the 2D Maxwell lattices. (a) The energy spectrum for
$M=2$; (b) The Berry curvature $F_{xy}(k_y)$ for $k_x=0$ and
$M=1.8,2,2.2$; (c) The energy spectrum for $M=0$; (d) The Berry
phase $\gamma$ as a function of the parameter $M$, which
corresponds to the Chern number $\mathcal{C}_1=\gamma/2\pi$ when the 2D system is in the insulating phase with $M\neq0,\pm2$. \label{2D-Bulk}}
\end{figure}

Under the periodic boundary condition, the model Hamiltonian
(\ref{2DHam}) can be  rewritten as
$\hat{H}=\sum_{\mathbf{k},ss'}\hat{a}^{\dag}_{\mathbf{k}s}[\mathcal{H}(\mathbf{k})]_{ss'}\hat{a}_{\mathbf{k}s'}$,
where
$\hat{a}_{\mathbf{k}s}=1/\sqrt{V}\sum_ke^{i\mathbf{k}\cdot\mathbf{r}}\hat{a}_{\mathbf{k}s}$
is the annihilation operator in momentum space
$\mathbf{k}=(k_x,k_y)$, and
$\mathcal{H}(\mathbf{k})=\mathbf{R}(\mathbf{k})\cdot
\mathbf{\hat{S}}$ is the Bloch Hamiltonian. Here
$\mathbf{R}(\mathbf{k})=(R_x,R_y,R_z)$ denotes the Bloch vectors:
$R_x=2t\sin{k_x}, R_y=2t\sin{k_y}$, and
$R_z=2t(M-\cos{k_x}-\cos{k_y})$, with the lattice spacing $a\equiv
1$ and $\hbar \equiv1$ hereafter. The energy spectrum of this
system is given by $E(\mathbf{k})=0,\pm|\mathbf{R}(\mathbf{k})|$,
which has a zero-energy flat band in the middle of the three
bands.

\emph{Maxwell quasiparticles in Maxwell metals.---} The three
bands $E(\mathbf{k})=0,\pm|\mathbf{R}(\mathbf{k})|$ touch at one
point when $M=\pm2$, and touch at two points when $M=0$. For the case $M=2$, the three bands touch at
$\mathbf{K}_+=(0,0)$ in the energy spectrum shown in Fig.
\ref{2D-Bulk}(a). We expand the Bloch Hamiltonian near this
threefold degenerate point and obtain the following effective
Hamiltonian for the low-energy excitations in the system
\begin{equation}
\mathcal{H}_{+}(\mathbf{q})=v\emph{q}_x\hat{S}_x+v\emph{q}_y\hat{S}_y,
\end{equation}
where $v=2t$ is the effective speed of light and
$\mathbf{q}=\mathbf{k}-\mathbf{K}_+$. This effective Hamiltonian
takes the Maxwell Hamiltonian $\hat{H}_{\text{M}}$ in two
dimensions and thus the dynamics of the low-energy excitations can
be effectively described by the Maxwell Hamiltonian
(\ref{MaxwellEq}). In this sense, we name these low-energy
excitations as the Maxwell quasiparticles and the threefold
degeneracy point as the Maxwell point. When the ultra-cold atoms
are fermions and the Fermi level lies near the Maxwell point, the
system can be named as the Maxwell metals, which is a metallic
state due to the existence of the zero-energy flat band.

To study the topological stability of the Maxwell point, we
consider the Berry phase for a Maxwell quasiparticle  circling
around the point $\gamma=\oint_c d\mathbf{k} \cdot
\mathbf{\mathbf{F}(k)}$, where the Berry curvature
$\mathbf{F(k)}=\nabla\times \mathbf{A(k)}$ with the Berry
connection defined by the wave function
$|\psi_n(\mathbf{k})\rangle$ in the $n$-th ($n=1,2,3$) band
$\mathbf{A(k)}=-i\langle\psi_n(\mathbf{k})|\nabla_{\mathbf{k}}|\psi_n(\mathbf{k})\rangle$.
For this three-band system described by the Bloch Hamiltonian
$\mathcal{H}(\mathbf{k})$, the lowest-band Berry curvature in the
$k_x$-$k_y$ space can be rewritten as \cite{He1}
\begin{equation}
F_{xy}=-\frac{1}{R^{3}}\mathbf{R}\cdot(\partial_{k_x}{\mathbf{R}}\times\partial_{k_y}{\mathbf{R}}).
\end{equation}
The distributions of $F_{xy}(k_y)$ for fixed $k_x=0$ and
typical parameters $M=1.8,2,2.2$ are plotted in Fig.
\ref{2D-Bulk}(b), and the results show that $F_{xy}$ is a
Dirac-$\delta$ function at the Maxwell point. The numerical
integration of $F_{xy}$ over the Brillouin zone for $M=2$ gives
the Berry phase $\gamma=2\pi$, which is confirmed by the
analytical calculation \cite{Supplement}.

For the case $M=-2$, the single Maxwell point moves to the
Brillouin edge $\mathbf{K}_-=(\pi,\pi)$ with the Berry phase
$\gamma=-2\pi$, and the low-energy effective Hamiltonian becomes
$\mathcal{H}_{-}(\mathbf{q})=-\mathcal{H}_{+}(\mathbf{q})$. When
$M=0$ with the energy spectrum shown in Fig. \ref{2D-Bulk}(c),
there are two Maxwell points at $(0,\pi)$ and $(\pi,0)$ with the
effective Hamiltonian $\mathcal{H}_0(\mathbf{q})=\pm
v\emph{q}_x\hat{S}_x\mp v\emph{q}_y\hat{S}_y$, respectively. In
this case, the Berry phase for both of the two Maxwell points is
obtained as $\gamma=0$, which corresponds to a trivial metallic
state.

\emph{Maxwell edge modes in Maxwell insulators.---} When the
parameter $M\neq0,\pm2$, the two band gaps are open and the system
becomes insulating states. Under this condition, we can calculate
the Chern number $\mathcal{C}_n$ for the corresponding three bands
with the band index $n$:
\begin{equation}
\mathcal{C}_n=\frac{1}{2\pi}\int_{BZ}{dk_xdk_y}
F_{xy}(k_x,k_y)=\gamma/2\pi.
\end{equation}
We find nonzero Chern numbers $\mathcal{C}_1=-\mathcal{C}_3=2\text{sign}(M)$ for
$|M|<2$ and $\mathcal{C}_1=\mathcal{C}_3=0$ for $|M|>2$
\cite{Supplement}, and thus $\mathcal{C}_2(M)=0$ for the flat band. Figure
\ref{2D-Bulk}(d) shows the Berry phase of the lowest band
$\gamma=2\pi \mathcal{C}_1$ as a function of the parameter $M$, which
indicates that this system is subjected to three topological phase
transitions at the tunable parameter $M=2,0,-2$.

\begin{figure}[htbp] \centering
\includegraphics[width=8.5cm]{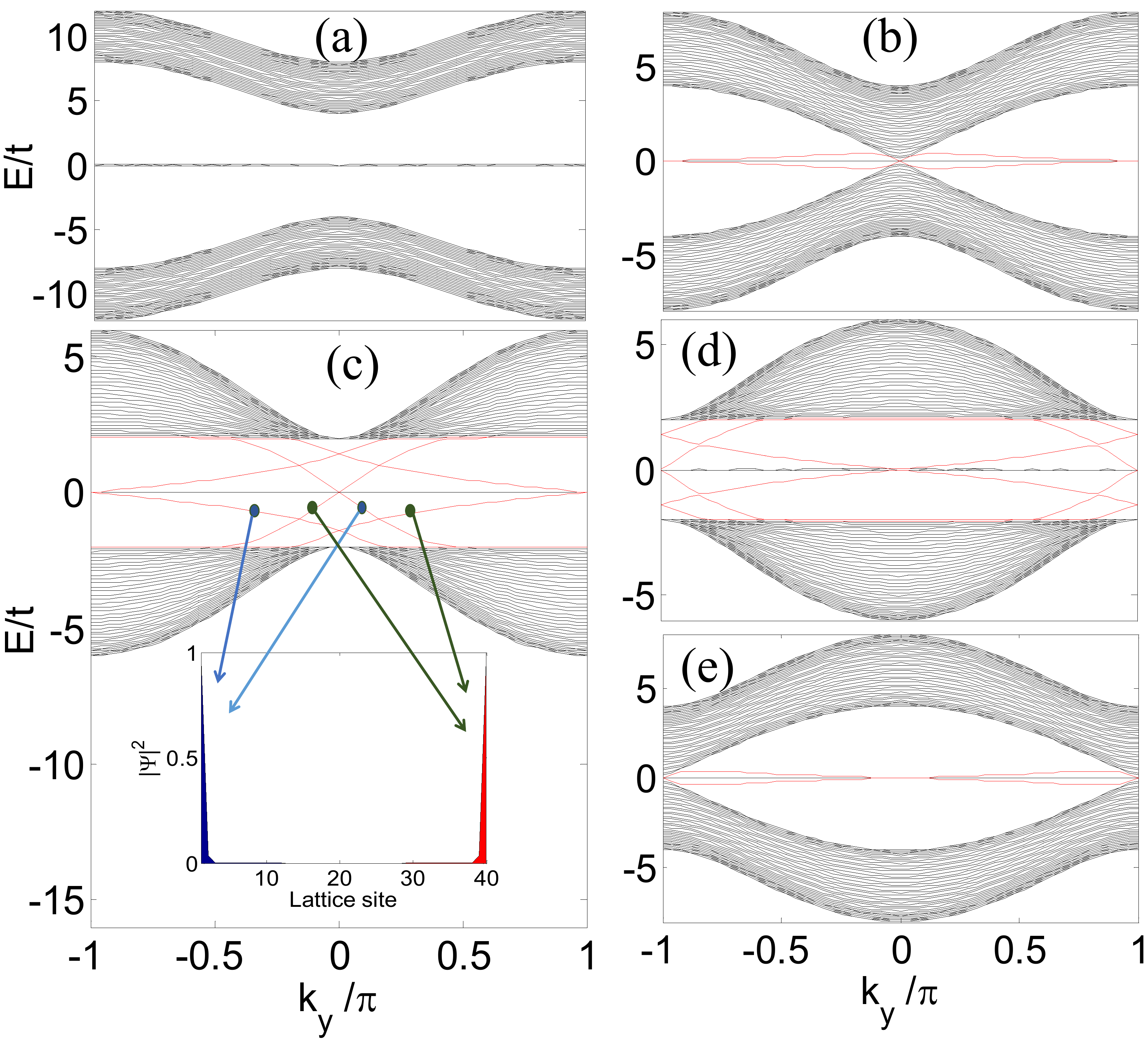}
 \caption{(color online). Energy spectra and edge states.  (a) $M=4$; (b)
$M=2$; (c) $M=1$; (d) $M=-1$; and (e) $M=-2$. The inse in (c)
shows the density distributions of four typical edge modes. The
edge modes in (a-e) are plotted in red. The lattice sites $L_x=40$
under open boundary condition.}\label{1D-Spectrum}
\end{figure}

To further study the topological properties of the system, we
numerically calculate the energy spectrum of a cylindrical surface
with periodic boundary condition for $y$ direction and the length
$L_x = 40$ under the open boundary condition along the $x$
direction, and the results are shown in Fig. \ref{1D-Spectrum}.
From Figs. \ref{1D-Spectrum}(a-e), we show the variation of the
energy spectra by changing the parameter $M$. For $M=4$ in Fig.
\ref{1D-Spectrum}(a), there is no edge mode in the two band
gaps in this trivial insulating state with the Chern number
$\mathcal{C}_n=0$. Decreasing the parameter $|M|$ at critical
values $M=\pm2$ [Figs. \ref{1D-Spectrum}(b) and
\ref{1D-Spectrum}(e)], the band gaps close and the system is in
the nontrivial Maxwell metallic phase with $\pm2\pi$ Berry phase
(corresponding to the Chern number $\pm1$) and a branch of edge
modes connecting the lowest (third) band and the middle flat band
\cite{Supplement}. For $M=\pm1$ [Figs. \ref{1D-Spectrum}(c) and
\ref{1D-Spectrum}(d)], the spectra contain two pairs asymmetric
branches of edge modes connecting the separated lowest (third)
band and the middle flat band, which is consistent with bulk-edge
correspondence in these cases with the bulk Chern number
$|\mathcal{C}_{1,3}|=2$. The density distributions of some edge
modes are shown in the inset in Fig. \ref{1D-Spectrum}(c) for
typical $k_y$.

\begin{figure}[htbp]\centering
\includegraphics[width=8.5cm]{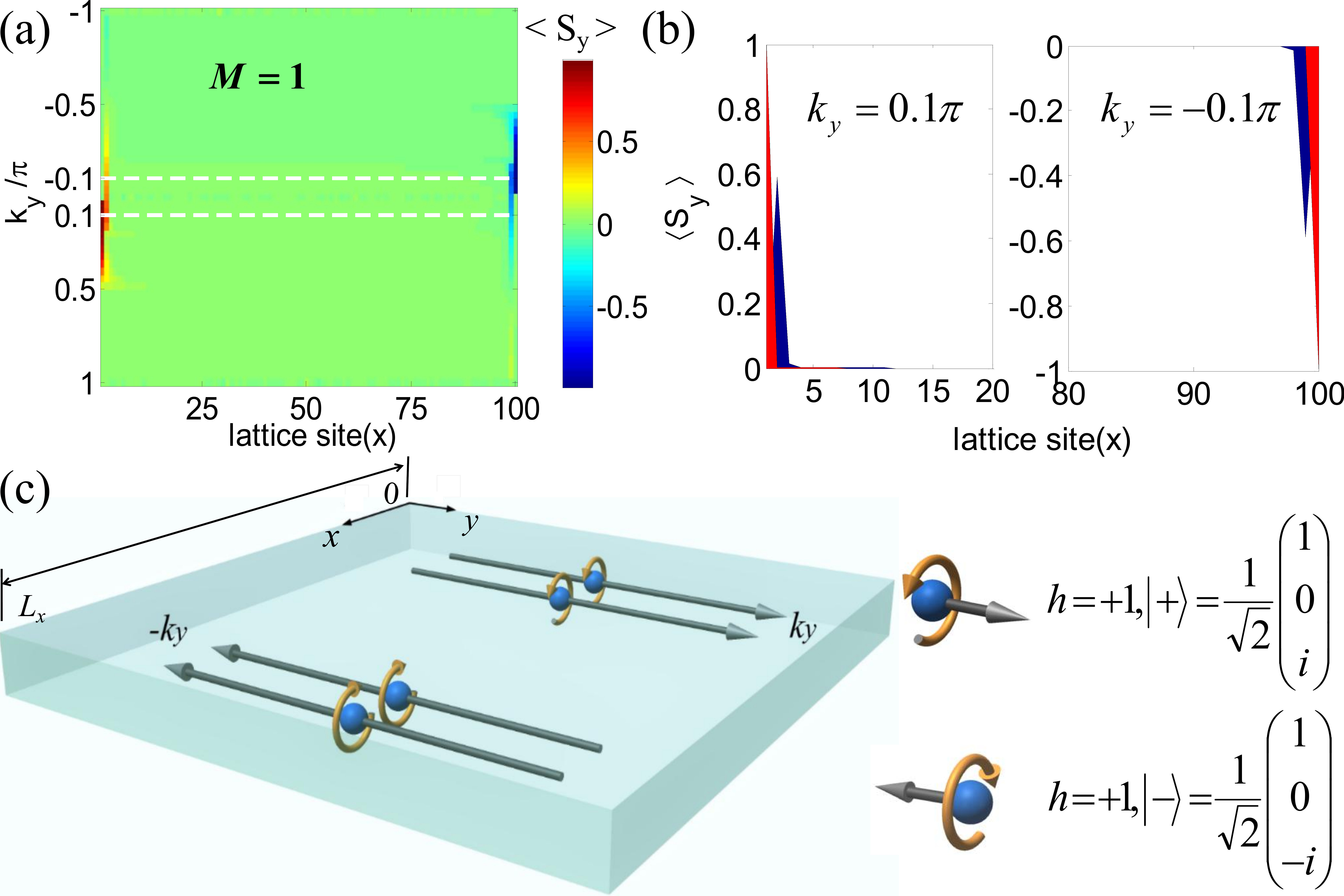}
 \caption{(color online). Maxwell quasiparticles as the edge states in the 2D topological Maxwell insulators.
 (a) Expectation value of $\hat{S}_y$ as a function of $k_y$ and $x$ for the
 reduced 1D chain with lattice sites $L_x=100$ under open boundary condition; (b) Density distribution of $\hat{S}_y(x)$ for $k_y=0.1\pi$  and $k_y=-0.1\pi$;
 (c) Schematic diagram for Maxwell edges states $|+\rangle$ and $|-\rangle$ in the Maxwell topological insulator with opposite momenta, both corresponding
  to the right circularly-polarized photons with the helicity $h=+1$.} \label{EdgeMode}
\end{figure}

Without loss of generality, we consider the topological insulating
state for $M=1$ and further explore the properties of the edge
modes in the first band gap. We find the correspondence
between the helicity of these edge states and the polarization of
photons, and thus named them as the Maxwell edge modes in this
so-called topological Maxwell insulator. In particular, we reveal
that this system exhibits an analogous quantum anomalous Hall
effect \cite{Nagaosa}, with the edge modes being strong
spin-momentum locking as eigenstates of the spin operator
$\hat{S}_y$. This means that the two bunches of quasiparticle
streams on the two edges can be treated as the streams of
polarized Maxwell quasiparticles moving along the $y$ axis.

In Fig. \ref{EdgeMode}(a), we numerically calculate the expectation
value $\langle\hat{S}_y\rangle$ with the wave functions of the 2D
lattice with $L_x=100$. The results show that the distribution of
$\langle\hat{S}_y(k_y,x)\rangle$ has two peaks localized at both
of the left and right edges with opposite signs. To be more
clearly, we plot $\langle\hat{S}_y(x)\rangle$ for $k_y=0.1\pi$ and
$k_y=-0.1\pi$ in Fig. \ref{EdgeMode}(b), respectively. It
indicates that just the two edge states for each edge are the
eigenstates of $\hat{S}_y$. The edge states on the $x=0$ edge with
positive eigenvalue are
$|+\rangle=\frac{1}{\sqrt{2}}\begin{pmatrix}1,0,i\end{pmatrix}^{T}=\frac{1}{\sqrt{2}}(\mathbf{e}_x+i\mathbf{e}_z)^T$,
and the ones on the $x=L_x$ edge with negative eigenvalue are
$|-\rangle=\frac{1}{\sqrt{2}}\left(\begin{matrix}1,0,-i\end{matrix}\right)^{T}=\frac{1}{\sqrt{2}}(\mathbf{e}_x-i\mathbf{e}_z)^T$,
where $\mathbf{e}_j$ ($j=x,y,z$) are the unit vectors of Cartesian
coordinates. Thus, the effective Hamiltonian of edge states is
given by
\begin{equation}
H_{\text{edge}}=v_yk_{y}\hat{S}_{y}.
\end{equation}
This effective Hamiltonian is nothing but the one-dimensioanl Maxwell
Hamiltonian of circularly-polarized quasiparticles. The helicity
operator is given by
$\hat{h}=\hat{\mathbf{S}}\cdot\frac{\mathbf{k}}{|\mathbf{k}|}=\text{sign}(k_y)\hat{S}_y$,
that is, the projection of the spin along the direction of the
linear momentum \cite{Lan,Greiner}, where
$\mathbf{k}=k_y\mathbf{e}_y$ denotes the momentum of Maxwell
particles moving along the $y$ direction. Thus, the edge
quasiparticle-streams in this Maxwell topological insulator can be
treated as Maxwell quasiparticle-streams with the same helicity
$h\equiv\langle \hat{h}\rangle=+1$ for opposite momenta, which
satisfies the helicity conservation of massless photons in quantum
field theory, as shown in Fig. \ref{EdgeMode}(c). In addition, the
momentum $\mathbf{k}$ can also be considered as the wave vector of
the plane electromagnetic wave propagated along the $y$ axis, both of the
edge states $|+\rangle$ $(k_y>0)$ and $|-\rangle$ $(k_y<0)$ with
same helicity can be regarded as right circularly-polarized
waves which consist of the two independent transverse polarization
vectors $\mathbf{e}_x$ and $\mathbf{e}_z$ with opposite momenta.
We can see from Fig. \ref{EdgeMode}(c) that the Maxwell edge modes
moving along $+y$ ($-y$) direction correspond to the right
circularly-polarized waves rotating anticlockwise (clockwise) in
the $xz$ plane (if one looks along the $-y$ axis) propagated along
$+y$ ($-y$) direction. Likewise, for the case $-2<M<0$ with the Chern number $\mathcal{C}=-2$, the
Maxwell-edge modes on the $x=0$ ($x=L_x$) edge with $h=-1$ are
corresponding to the left circularly-polarized waves propagated
along $-y$ ($+y$) direction similar to the schematic diagram in
Fig. \ref{EdgeMode}(c). Because the electromagnetic waves are
transverse waves, there is no longitudinal component and thus no
edge mode with the helicity $h=0$, which corresponds to the unit
wave vector of plane waves. So, the Maxwell edge modes here with
such strong spin-momentum locking perfectly correspond to the
circularly-polarized photons.

\emph{Discussions and conclusion.---} We now address some possible
concerns on the realistic cold-atom experiments. In our
proposed OL scheme for implementing the model
Hamiltonian (\ref{2DHam}), the major difficulty is to realize the
spin-flip hopping terms along each direction, which will involves
a number of Raman beams \cite{Supplement}. However, all of the
lasers can be drawn from the same one with the small relative
frequency shift induced by an acoustic optical modulator. The
Maxwell points in the band structures with the related topological
phase transition can be detected from the Bragg spectroscopy or
Bloch-Zener oscillations, similar to the methods for detecting
Dirac point in OLs \cite{Zhu,Tarruell}. In addition,
the Berry curvature and thus the Chern numbers (Berry phases) can
be measured by the newly-developed technique of tomography of Bloch band in OLs \cite{BandTomography,Li}, and the Chern
numbers can also be revealed from the shift of the hybrid Wannier
center of an atomic cloud \cite{Supplement,Lohse,LuBEC,Nakajima}.

In summary, we have explored the topological properties of Maxwell
quasiparticles emerged in Maxwell metals and Maxwell insulators.
The proposed model could be realized in OLs and the
exotic properties of these topological quasiparticles can be
detected in cold-atom experiments. Moreover, the properties of the
topological Maxwell quasiparticles analogous to the Dirac and Weyl
fermions can be further investigated, such as wave dynamics of the
Maxwell fermions by studying the relativistic Zitterbewegung
oscillations \cite{Vaishnav,ZLi} or Klein tunneling
\cite{ZhangPRA2012} in the Maxwell insulators, the transport
properties of the Maxwell fermions and so forth.

\acknowledgments We thank Hai-Jun Zhang and  Rui-Bin Liu for
useful discussions. This work was supported by the NKRDP of China
(Grant No. 2016YFA0301803), the NSFC (Grants No. 11474153 and
11604103), and the PCSIRT (Grant No. IRT1243). D.-W. Z. was also
supported by the NSF of Guangdong Province (Grant No.
2016A030313436) and the Startup Foundation of SCNU.

\onecolumngrid
\section{Supplemental Material: Maxwell Quasiparticles Emerged in Optical Lattices}

\begin{quote}
In this supplemental materials, we provide more details on the
derivation of the Maxwell equations in the form of the
Schr\"{o}dinger equation, the schemes to realize the Maxwell
Hamiltonian with cold atoms in 2D optical lattices, the
derivation of the topological invariants, and a scheme to
measure the topological Chern numbers with cold atoms.
\end{quote}

\renewcommand{\thesection}{S.\Roman{section}}
\renewcommand{\thefigure}{S\arabic{figure}}

\section{Maxwell equations in the form of the Schr\"{o}dinger equation}

The well-known Maxwell equations in a vacuum are given by
\begin{equation}
\tag{S.1}
\begin{split}
\frac{1}{c}\frac{\partial{\mathbf{\tilde{E}}}}{\partial{t}}=\nabla\times\mathbf{\tilde{H}},\quad\quad
\nabla\cdot\mathbf{\tilde{E}}=0,\\
-\frac{1}{c}\frac{\partial{\mathbf{\tilde{H}}}}{\partial{t}}=\nabla\times\mathbf{\tilde{E}},\quad\quad
\nabla\cdot\mathbf{\tilde{H}}=0,
\end{split}
\end{equation}
where $\mathbf{\tilde{E}}=\sqrt{\varepsilon_0}\mathbf{E}$ with
$\mathbf{E}$ being the electric field,
$\mathbf{\tilde{H}}=\sqrt{\mu_0}\mathbf{H}$ with $\mathbf{H}$
being the magnetizing field, $c=1/\sqrt{\varepsilon_0\mu_0}$  is
the speed of light in the vacuum, and
$\varepsilon_0$, $\mu_0$ are the permeability and permittivity of the vacuum, respectively. If we
define the photon wave function as
$\mathbf{\Phi}(\mathbf{r},t)=\mathbf{\tilde{E}}(\mathbf{r},t)+i\mathbf{\tilde{H}}(\mathbf{r},t)$
\cite{Oppenheimer,Good} , we have $\nabla\cdot\mathbf{\Phi}=0$ and
$\nabla\times\mathbf{\Phi}=\nabla\times\mathbf{\tilde{E}}+i\nabla\times\mathbf{\tilde{H}}=\frac{i}{c}\frac{\partial{\mathbf{\Phi}}}{\partial{t}}$.
Thus we can obtain
\begin{equation}
\tag{S.2}
\begin{split}
\varepsilon_{\alpha\beta\gamma}\frac{\partial\Phi_m^{\gamma}}{\partial\beta}=
\frac{i}{c}\frac{\partial{\Phi_m^{\alpha}}}{\partial{t}}\Rightarrow{ci\varepsilon_{\alpha\beta\gamma}\hat{P}_{\beta}\Phi}_m^{\gamma}
=i\hbar\frac{\partial{\Phi_{m}^{\alpha}}}{\partial{t}},
\end{split}
\end{equation}
where $\hat{P}_{\beta}=-i\hbar\partial_{\beta}$. Let
$\hat{S}_{\alpha\gamma}^{\beta}=i\varepsilon_{\alpha\beta\gamma}$
with $\varepsilon_{\alpha\beta\gamma}$ $(\alpha, \beta, \gamma=x,
y, z)$ being the Levi-Civita symbol. We can rewrite the Maxwell
Eqs (S.1) as
\begin{equation}
\tag{S.3}
\begin{split}
i\hbar\frac{\partial}{\partial{t}}\mathbf{\Phi}=c\hat{\mathbf{S}}\cdot\hat{\mathbf{P}}\mathbf{\Phi},
\end{split}
\end{equation}
where the spin matrices are defined as
\begin{equation}
\tag{S.4}
\begin{split}
\hat{S}_x=\hat{S}^{1}= i\left(\begin{matrix}
    \varepsilon_{111}&\varepsilon_{112}&\varepsilon_{113}\\
    \varepsilon_{211}&\varepsilon_{212}&\varepsilon_{213}\\
    \varepsilon_{311}&\varepsilon_{312}&\varepsilon_{313}
    \end{matrix}
   \right)=
\left(\begin{matrix}
0&0&0\\
0&0&-i\\
0&i&0
\end{matrix}
\right),\\
\hat{S}_y=\hat{S}^{2}= i\left(\begin{matrix}
    \varepsilon_{121}&\varepsilon_{122}&\varepsilon_{123}\\
    \varepsilon_{221}&\varepsilon_{222}&\varepsilon_{223}\\
    \varepsilon_{321}&\varepsilon_{322}&\varepsilon_{323}
    \end{matrix}
   \right)=
\left(\begin{matrix}
    0&0&i\\
    0&0&0\\
    -i&0&0
    \end{matrix}
\right),\\
\hat{S}_z=\hat{S}^{3}= i\left(\begin{matrix}
    \varepsilon_{131}&\varepsilon_{132}&\varepsilon_{133}\\
    \varepsilon_{231}&\varepsilon_{232}&\varepsilon_{233}\\
    \varepsilon_{331}&\varepsilon_{332}&\varepsilon_{333}
    \end{matrix}
   \right)=
\left(\begin{matrix}
    0&-i&0\\
    i&0&0\\
    0&0&0
    \end{matrix}
\right).
\end{split}
\end{equation}
One can check that $[\hat{S}_x,\hat{S}_y]=i\hat{S}_z$,
$\mathbf{\hat{S}}\times\mathbf{\hat{S}}=i\mathbf{\hat{S}}$,
$S^2=S_x^2+S_y^2+S_z^2=2\begin{pmatrix}1&0&0\\0&1&0\\0&0&1\end{pmatrix}=S(S+1)$,
and $S=1$. Thus we obtain the Maxwell equations in the form of the
Schr\"{o}dinger equation, and the related Hamiltonian of single
photon with spin-1 is
$\hat{H}=c\hat{\mathbf{S}}\cdot\hat{\mathbf{P}}$. We can easily
obtain the eigenstates of $\hat{S}_y$, which are given by
$\hat{S}_y\Phi=s_y\Phi$ and
\begin{equation}
\tag{S.5}
\begin{aligned}
\Phi_{1}&=&\frac{1}{\sqrt{2}}\begin{pmatrix}1\\0\\i\end{pmatrix};&&s_y&=1\\
\Phi_{0}&=&\begin{pmatrix}0\\1\\0\end{pmatrix}; &&s_y&=0\\
\Phi_{-1}&=&\frac{1}{\sqrt{2}}\begin{pmatrix}1\\0\\-i\end{pmatrix};&&s_y&=-1\\
\end{aligned}
\end{equation}

The above derivation can be generalized to an anisotropic
medium. In a region without free charges and currents, the Maxwell
equations are given by
\begin{equation}\tag{S.6}
\begin{split}
\nabla\times\mathbf{E}&=-\frac{\partial{\mathbf{B}}}{\partial{t}},\quad
\nabla\cdot\mathbf{{E}}=0,\\
\nabla\times\mathbf{H}&=\ \
\frac{\partial{\mathbf{D}}}{\partial{t}},\quad\
\nabla\cdot\mathbf{B}=0,
\end{split}
\end{equation}
where the displacement field
$\mathbf{D}=\varepsilon_0\varepsilon_r\mathbf{E}$,
the magnetic field $\mathbf{B}=\mu_0\mu_r\mathbf{H}$, $\varepsilon_r$ and $\mu_r$ are
the relative permittivity and permeability, respectively. In the
anisotropic medium, $\varepsilon_r$ and $\mu_r$ become tensors
rather than numbers. To simplify the proceeding analysis, we
assume that the tensors $\varepsilon_r$ and $\mu_r$ are
simultaneously diagonalized, i.e.,
\begin{equation}\tag{S.7}
\begin{split}
\varepsilon_r=\left(\begin{matrix}
\varepsilon_x&0&0\\
0&\varepsilon_y&0\\
0&0&\varepsilon_z
\end{matrix}
\right),\ \mu_r=\left(\begin{matrix}
\mu_x&0&0\\
0&\mu_y&0\\
0&0&\mu_z
\end{matrix}
\right).
\end{split}
\end{equation}
The relationships between $\mathbf{D}$ and $\mathbf{E}$,
$\mathbf{B}$ and $\mathbf{H}$ now become
\begin{equation}\tag{S.8}
\begin{split}
\begin{pmatrix}D_x\\D_y\\D_z\end{pmatrix}=\varepsilon_0\begin{pmatrix}\varepsilon_x&0&0\\0&\varepsilon_y&0\\0&0&\varepsilon_z\end{pmatrix}\begin{pmatrix}E_x\\E_y\\E_z\end{pmatrix},\
\begin{pmatrix}B_x\\B_y\\B_z\end{pmatrix}=\mu_0\begin{pmatrix}\mu_x&0&0\\0&\mu_y&0\\0&0&\mu_z\end{pmatrix}\begin{pmatrix}H_x\\H_y\\H_z\end{pmatrix}.
\end{split}
\end{equation}
Thus Eq. (S.6) can be rewritten as
\begin{equation}\tag{S.9}
\begin{aligned}
\nabla\times\mathbf{E}&=-\frac{\partial{\mathbf{B}}}{\partial{t}}&\Rightarrow\epsilon_{\alpha\beta\gamma}\frac{\partial{E_\gamma}}{\partial\beta}
=-\frac{\partial{B_\alpha}}{\partial{t}}
&\Rightarrow\frac{c}{\sqrt{\varepsilon_\gamma\mu_\alpha}}\epsilon_{\alpha\beta\gamma}\frac{\partial{\tilde{E}_\gamma}}{\partial\beta}
=-\frac{\partial{\tilde{H}_\alpha}}{\partial{t}},\\
\nabla\times\mathbf{H}&=\ \
\frac{\partial{\mathbf{D}}}{\partial{t}}&\Rightarrow\epsilon_{\alpha\beta\gamma}\frac{\partial{H_\gamma}}{\partial\beta}=\
\frac{\partial{D_\alpha}}{\partial{t}}
&\Rightarrow\frac{c}{\sqrt{\varepsilon_\alpha\mu_\gamma}}\epsilon_{\alpha\beta\gamma}\frac{\partial{\tilde{H}_\gamma}}{\partial\beta}=\
\frac{\partial{\tilde{E}_\alpha}}{\partial{t}},
\end{aligned}
\end{equation}
where
$\tilde{E}_\alpha=\sqrt{\varepsilon_0\varepsilon_\alpha}E_\alpha$,
$\tilde{H}_\alpha=\sqrt{\mu_0\mu_\alpha}H_\alpha$. Then we define
the  photon wave function as
$\mathbf{\Phi}(\mathbf{r},t)=\mathbf{\tilde{E}}(\mathbf{r},t)+i\mathbf{\tilde{H}}(\mathbf{r},t)$,
we have $\nabla\cdot\mathbf{\Phi}=0$, and
\begin{equation}\tag{S.10}
\begin{split}
i\hbar\frac{\partial{\Phi_m^\alpha}}{\partial{t}}=\frac{c}{\sqrt{\varepsilon_\gamma\mu_\alpha}}(i\epsilon_{\alpha\beta\gamma})\frac{\hbar}{i}\frac{\partial{\tilde{E}_\gamma}}{\partial\beta}
+i\frac{c}{\sqrt{\varepsilon_\alpha\mu_\gamma}}(i\epsilon_{\alpha\beta\gamma})\frac{\hbar}{i}\frac{\partial{\tilde{H}_\gamma}}{\partial\beta}.
\end{split}
\end{equation}
Let $\nu_{\alpha\gamma}=c/\sqrt{\varepsilon_\alpha\mu_\gamma}$,
$\nu_{\gamma\alpha}=c/\sqrt{\varepsilon_\gamma\mu_\alpha}$,
$\hat{P}_\beta=-i\hbar\partial_{\beta}$,  when
$\varepsilon_\alpha\mu_\gamma=\varepsilon_\gamma\mu_\alpha$, that
is, $\nu_{\alpha\gamma}=\nu_{\gamma\alpha}$ [the condition for
obtaining a hermitian Hamiltonian, see Eq. (S.12)], then we
can further rewrite Eq. (S.10) as
\begin{equation}\tag{S.11}
\begin{split}
i\hbar\frac{\partial{\Phi_m^\alpha}}{\partial{t}}=\nu_{\alpha\gamma}(i\epsilon_{\alpha\beta\gamma})\hat{P}_\beta\frac{\partial{\Phi_m^\gamma}}{\partial\beta}.
\end{split}
\end{equation}
Thus we obtain the following Schr\"{o}dinger's equation
\begin{equation}\tag{S.12}
\begin{split}
i\hbar\frac{\partial}{\partial{t}}\begin{pmatrix}\Phi_m^x\\
\Phi_m^y\\ \Phi_m^z\end{pmatrix}=
\begin{pmatrix}0&-i\nu_{xy}\hat{P}_z&i\nu_{xz}\hat{P}_y\\i\nu_{yx}\hat{P}_z&0&-i\nu_{yz}\hat{P}_x\\ -i\nu_{zx}\hat{P}_y&i\nu_{zy}\hat{P}_x&0\end{pmatrix}
\begin{pmatrix}\Phi_m^x\\ \Phi_m^y\\ \Phi_m^z\end{pmatrix}.
\end{split}
\end{equation}
This corresponds to the Maxwell equations in the anisotropic medium in the Schr\"{o}dinger's form
$$i\hbar\frac{\partial}{\partial{t}}\mathbf{\Phi}=\hat{H}\mathbf{\Phi},$$
where the Hamiltonian is given by
$$\hat{H}=v_x\hat{S}_x\hat{P}_x+v_y\hat{S}_y\hat{P}_y+v_z\hat{S}_z\hat{P}_z.$$
Here
$\hat{S}_\beta=(\hat{S}_{\alpha\gamma})^\beta=i\epsilon_{\alpha\beta\gamma}$
has the same form as that in Eq. (S.4). Noted that,
$v_x=\nu_{yz}=\nu_{zy}$, $v_y=\nu_{zx}=\nu_{xz}$, and
$v_z=\nu_{xy}=\nu_{yx}$ are the necessary and sufficient condition
to obtain a hermitian Hamiltonian in Eq. (S12).
 It returns to the
free space situation in Eq. (S.3) when $\varepsilon_r=\mu_r=1$.

\section{Realizing the 2D model Hamiltonian in optical lattices}

In this section, we propose two different schemes to realize the
Maxwell quasi-particles in the 2D optical lattices.

In the first part, we provide some details on the realization
scheme based on the Raman-assisted tunneling method
\cite{LAT1,LAT2,LAT3} to implement the following 2D model
Hamiltonian of Maxwell insulators and Maxwell metals:
\begin{equation} \tag{S.13}
\begin{split}
\hat{H}_{2D}=&t\sum_{\mathbf{r}}\left[ \hat{H}_{\mathbf{rx}}+\hat{H}_{\mathbf{ry}}+ \left( \Gamma_0\hat{a}^{\dag}_{\mathbf{r},0}\hat{a}_{\mathbf{r},\uparrow} +\textrm{H.c.}\right) \right],\\
\hat{H}_{\mathbf{rx}}=&-\hat{a}^{\dag}_{\mathbf{r-x},0}(\hat{a}_{\mathbf{r},\downarrow}+i\hat{a}_{\mathbf{r},\uparrow}) + \hat{a}^{\dag}_{\mathbf{r+x},0}(\hat{a}_{\mathbf{r},\downarrow}-i\hat{a}_{\mathbf{r},\uparrow})+\textrm{H.c.},\\
\hat{H}_{\mathbf{ry}}=&\hat{a}^{\dag}_{\mathbf{r-y},\uparrow}(\hat{a}_{\mathbf{r},\downarrow}+i\hat{a}_{\mathbf{r},0})
-
\hat{a}^{\dag}_{\mathbf{r+y},\uparrow}(\hat{a}_{\mathbf{r},\downarrow}-i\hat{a}_{\mathbf{r},0})+\textrm{H.c.}.
\end{split} \label{2DHam}
\end{equation}
\begin{figure}[htbp]\centering

\includegraphics[width=12cm]{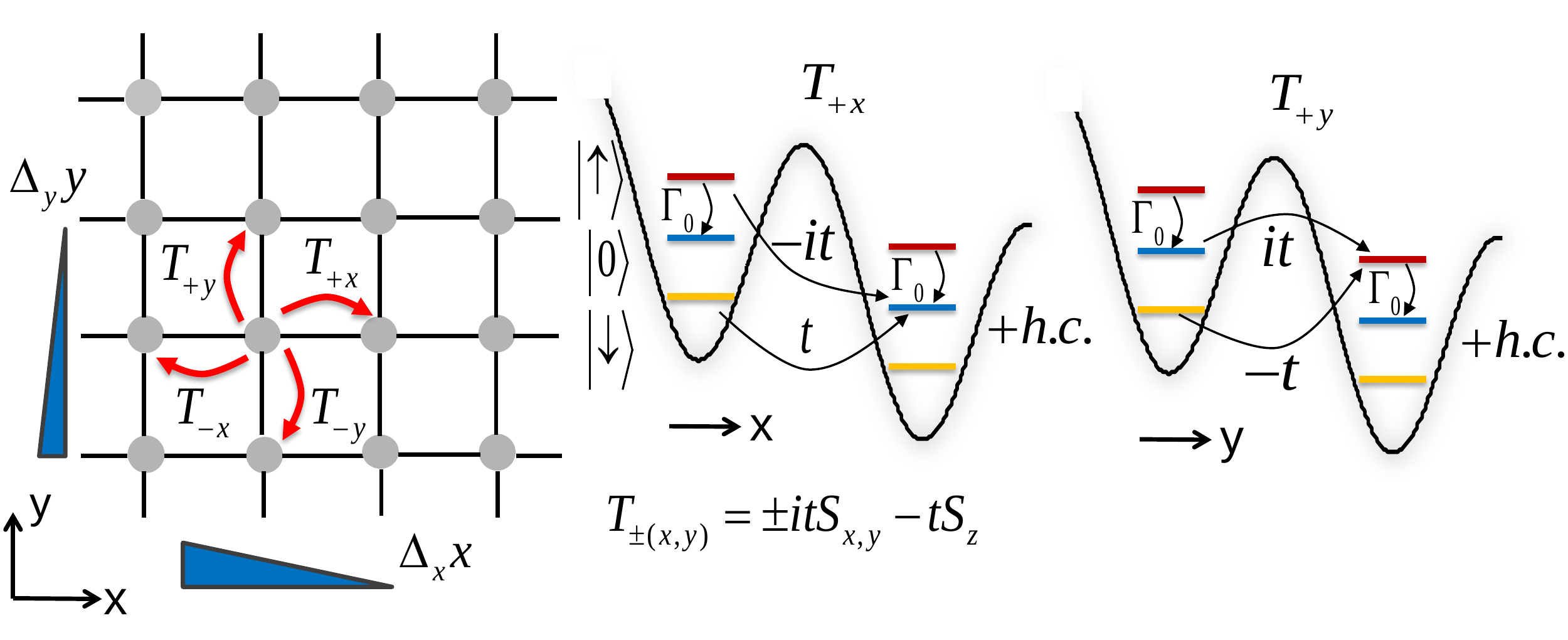}
 \caption{(Color online) Schematic diagram of realizing the model Hamiltonian (\ref{2DHam}) with the spin-flip hopping and on-site spin-flip terms
 with cold fermionic atoms in a tilted square optical lattice. The three atomic internal states $|\uparrow\rangle,|0\rangle,|\downarrow\rangle$ form the (pseudo)spin-1 basis and the required atomic spin-flip hopping, such as $T_{+x}$ and $T_{+y}$, are shown in the figure. Along the $x$ ($y$) axis, the natural hopping is  suppressed due to a large tilt potential $\Delta_xx$ ($\Delta_yy$) with the energy difference $\Delta_x$ ($\Delta_y$) per lattice site. The required hopping can then be restored and engineered by using proper Raman lasers. } \label{2dOL}
\end{figure}
We can use noninteracting fermionic (or bosonic) atoms in a titled square optical lattice and choose three atomic internal states in the ground state manifold to encode the three spin states $|s\rangle$ ($s=\uparrow,0,\downarrow$), as shown in Fig. \ref{2dOL}. The other levels in the ground state manifold are irrelevant because they can be depopulated by the optical pumping and transition. Here, the on-site spin-flip term $\Gamma_0\hat{a}^{\dag}_{\mathbf{r},0}\hat{a}_{\mathbf{r},\uparrow}$ can be easily achieved by applying a simple radio-frequency field or Raman beams for coupling the atomic internal states. Thus, the major difficulty for implementing this model Hamiltonian is to realize the spin-flip hopping terms $\hat{H}_{\mathbf{rx}}$ and $\hat{H}_{\mathbf{ry}}$ along each direction shown in Fig. \ref{2dOL}. Here the spin-flip hopping terms can be diagrammatically visualized as
\begin{align}
x& \text{-}\text{direction:~} T_{+x}+T_{-x}=\overset{\times }{
\curvearrowleft }|1_x\rangle\overset{\sqrt{2}}{\curvearrowright }|0\rangle+|0\rangle\overset{-\sqrt{2}}{\curvearrowleft}%
|2_x\rangle\overset{\times }{\curvearrowright }+\text{
H.c.}, \notag\\
y& \text{-}\text{direction:~} T_{+y}+T_{-y}=\overset{\times }{
\curvearrowleft }|1_y\rangle\overset{-\sqrt{2}}{\curvearrowright }|\uparrow\rangle+|\uparrow\rangle\overset{\sqrt{2}}{\curvearrowleft}%
|2_y\rangle\overset{\times }{\curvearrowright }+\text{
H.c.}, \notag
\end{align}
where $\overset{\times }{\curvearrowright }$ indicates that the hopping is
forbidden along this direction, and the states $|1_x\rangle=\left(|\downarrow\rangle-i|\uparrow\rangle\right)/\sqrt{2}$, $|2_x\rangle=\left(|\downarrow\rangle+i|\uparrow\rangle\right)/\sqrt{2}$, $|1_y\rangle=\left(|\downarrow\rangle-i|0\rangle\right)/\sqrt{2}$, $|2_y\rangle=\left(|\downarrow\rangle+i|0\rangle\right)/\sqrt{2}$ are superpositions of the original spin-basis vectors $|\uparrow\rangle$, $|0\rangle$, $|\downarrow\rangle$.

We can use Raman-assisted tunneling \cite{LAT1,LAT2,LAT3} to realize the spin-flip hopping terms depicted above. First, the required broken parity (left-right) symmetry in these hopping terms can be achieved by titling the square optical lattice with a homogeneous energy gradient
along the $x$ and $y$-directions. This can be realized through the natural gravitational field or the gradient of a dc- or ac-Stark shift. Note that the Raman-assisted hopping in tilted optical lattices has been demonstrated in recent experiments \cite{LAT2}. In our scheme, we require different
linear energy shifts per site $\Delta_{x,y}$ along the $x$ and $y$-directions as shown in Fig. \ref{2dOL}, such as $\Delta _{x}\approx 1.5\Delta _{y}$. Secondly, the natural hopping is suppressed by the large tilt potential $\Delta _{y}\gg t_{0}$ with $t_0$ denoting the natural tunneling rate. Under this condition, the hopping probability $\left(t_{0}/\Delta _{y}\right) ^{2}$ induced by the natural tunneling is negligible in this tilted lattice. Finally, the hopping terms can be restored and engineered by application of two-photon Raman transitions with the laser beams of proper configurations.

\begin{figure}[htbp]\centering
\includegraphics[width=10cm]{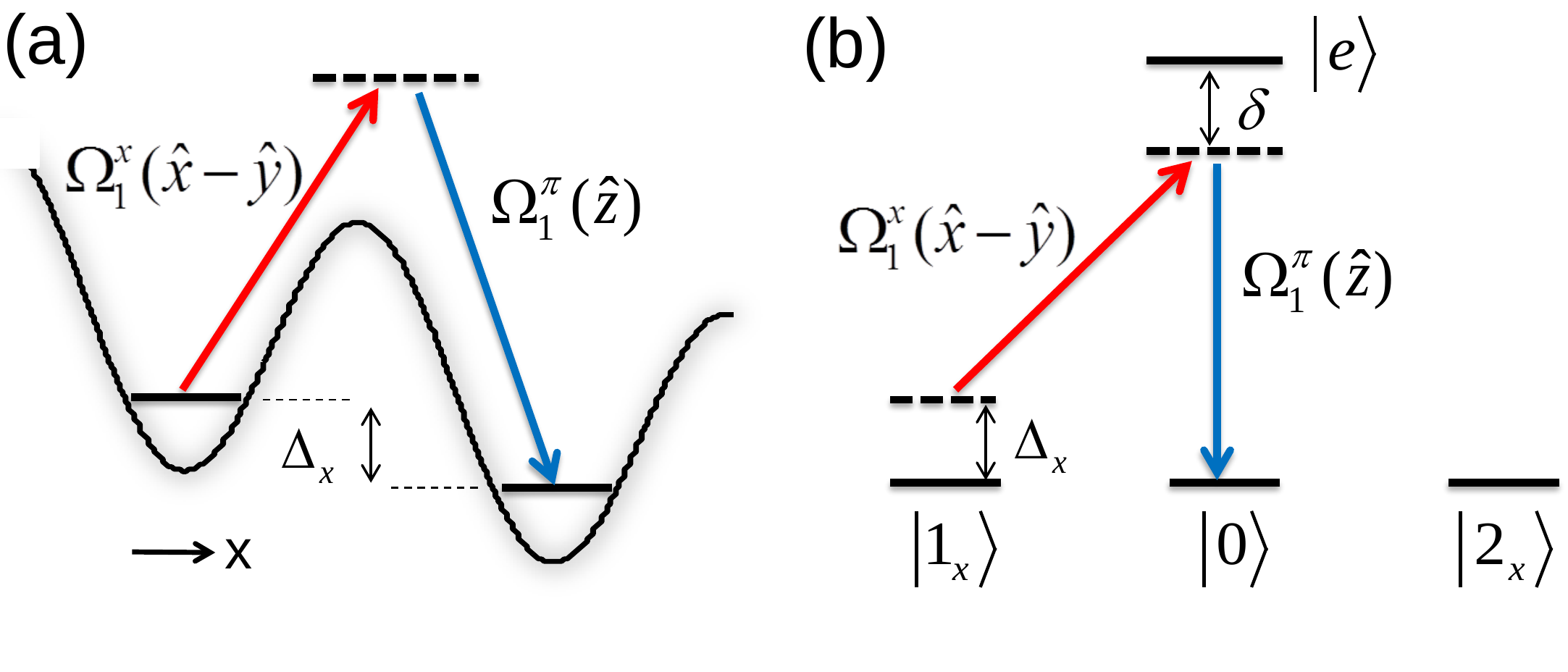}
\caption{(Color online) (a) A linear tilt $\Delta_{x}$ per lattice site along $x$-direction and the Raman beams $\Omega _{1}^{x}$ and $\Omega _{1}^{\pi}$ for addressing atoms. (b) The two Raman beams used to induce the required atomic tunneling term $T^{(1)}_{+x}$. The vector units in brackets show the polarization direction of the corresponding beam.} \label{x-hopping}
\end{figure}

Let us first consider a single term $T_{+x}^{(1)}=\hat{a}^{\dag}_{\mathbf{r+x},0}(\hat{a}_{\mathbf{r},\downarrow}-i\hat{a}_{\mathbf{r},\uparrow})$ to explain our Raman-assisted hopping scheme. This corresponds to an atom in the spin state $|1_x\rangle=\left(|\downarrow\rangle-i|\uparrow\rangle\right)/\sqrt{2}$ at site $\mathbf{r}$ hopping to site $\mathbf{r+x}$ while changing the spin state to $|0\rangle$ with hopping strength $\sqrt{2}$, which can be diagrammatically visualized as
\begin{equation}
T_{+x}^{(1)}=\hat{a}^{\dag}_{\mathbf{r+x},0}(\hat{a}_{\mathbf{r},\downarrow}-i\hat{a}_{\mathbf{r},\uparrow}) \qquad \Longleftrightarrow \qquad x\text{-}\text{direction:}\quad \overset{\times }{
\curvearrowleft }|1_x\rangle\overset{\sqrt{2}}{\curvearrowright }|0\rangle. \notag
\end{equation}
This hopping term can be achieved by two Raman beams $\Omega _{1}^{x}(\hat{x}-\hat{y})=\sqrt{2}\Omega _{0}e^{ikz}$ polarized along $(\hat{x}-\hat{y})$-direction and $\Omega _{1}^{\pi}(\hat{z})=\Omega _{0}e^{ikx}$ with $\pi$-polarization along $\hat{z}$-direction, as shown in Fig. \ref{x-hopping}. Here the population of the excited state $|e\rangle$ which is estimated by $|\Omega_{0}/\delta|^{2}$ is negligible due to the large single-photon detuning $\delta$. The two-photon detuning $\Delta_{x}$ matching the linear energy shift of the lattice per site ensures that it only allows $|1_x\rangle$ hopping to the right, and the other direction is forbidden by a large energy mismatch $2\Delta_{x}$. We can address the spin states through the polarization selection rule since the original spin basis $|\downarrow\rangle,|0\rangle, |\uparrow\rangle$ differ in the magnetic quantum number by one successively. Thus, a $\pi$-polarized beam $\Omega_{1}^{\pi}$ excites the state $|0\rangle$ and a linear $(\hat{x}-\hat{y})$-polarized beam $\Omega_{1}^{x}$ excites the superposition state $|1_x\rangle=\left(|\downarrow\rangle-i|\uparrow\rangle\right)/\sqrt{2}$ as the polarization $(\hat{x}-\hat{y}) \sim (\sigma^{+}-i\sigma^{-})$. These two beams together induce a Raman-assisted hopping between $|1_x\rangle$ and $|0\rangle$. The hopping amplitude and phase are controlled by the corresponding Raman beam amplitude and phase \cite{LAT1,LAT2,LAT3}, which can be written as
\begin{equation} \tag{S.14}
t_{\mathbf{r,+x}}=\frac{\sqrt{2}|\Omega_0|^2}{\delta}\beta e^{i\delta
\mathbf{k}\cdot\mathbf{r}},~~\beta=\int dxw^{\ast
}(x+a)e^{-ikx}w(x)\int dyw^{\ast }(y)w(y).
\end{equation}
Here $\delta \mathbf{k}=(-k,0)$ for this hopping term and we have used factorization of the Wannier function $w(\mathbf{r}^{\prime
})=w(x^{\prime })w(y^{\prime })$ in a square lattice. If we adjust the interfering angle of the lattice beams to satisfy the condition $ka=2\pi$, the site dependent phase term can always be reduced to $e^{i\delta \mathbf{k}\cdot \mathbf{r}}=1$. Under this condition, we can obtain the required hopping strength $t_1=\sqrt{2}t$ with $t=\beta|\Omega_0|^2/\delta$.

\begin{figure}[htbp]\centering
\includegraphics[width=10cm]{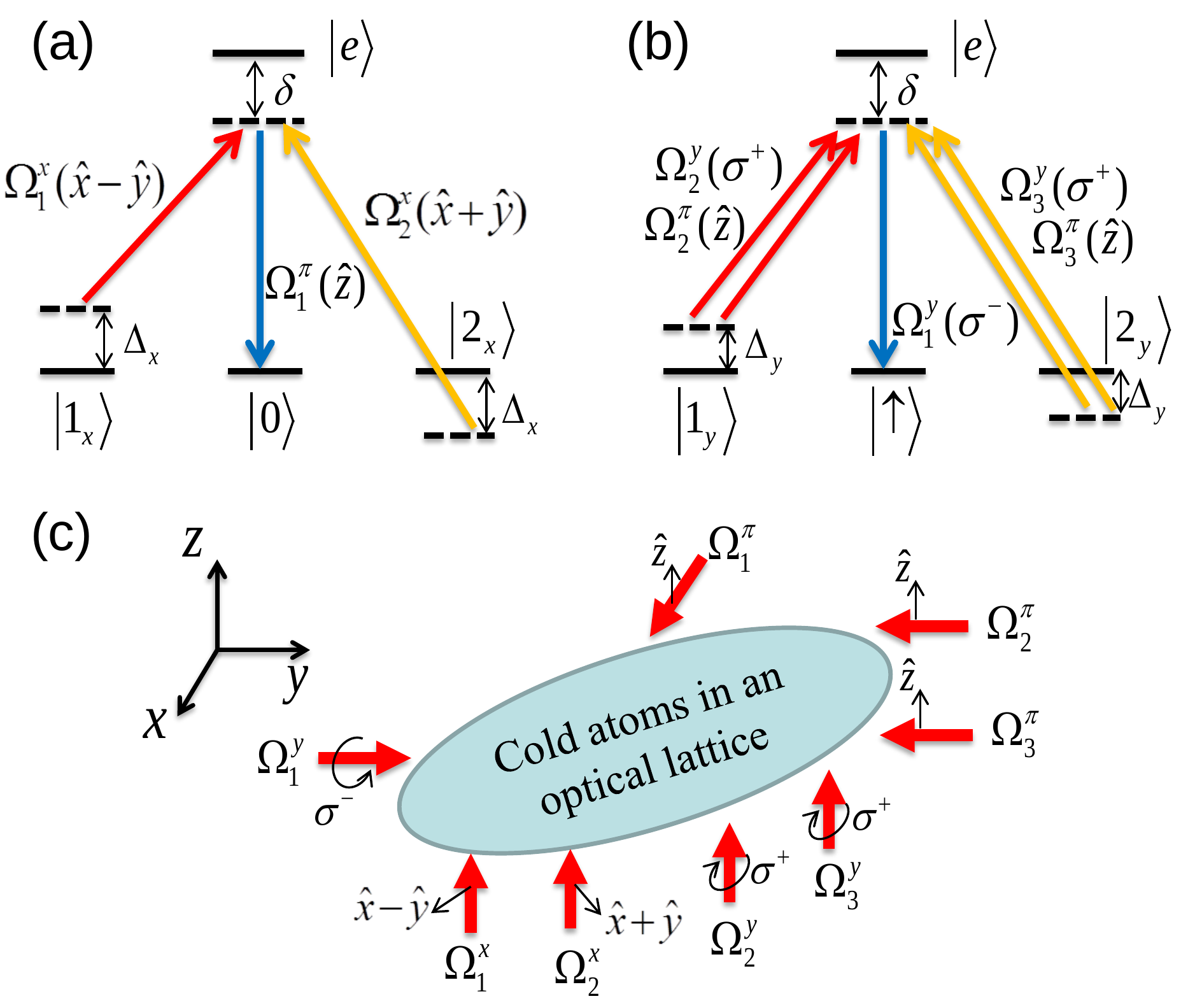}
\caption{(Color online) Schematics of the laser configuration to realize the Hamiltonian. (a) The three Raman beams for inducing the desired hopping along $x$ axis; (b) The five Raman beams for inducing the desired hopping along $y$ axis; (c) The total lasers with the corresponding polarization and propagation direction. The detuning in each direction matches the frequency offset of the corresponding Raman beams.} \label{RamanBeam}
\end{figure}

All the other hopping terms in the model Hamiltonian can be realized in a similar manner. For examples, the hopping term $T_{-x}^{(2)}=-\hat{a}^{\dag}_{\mathbf{r-x},0}(\hat{a}_{\mathbf{r},\downarrow}+i\hat{a}_{\mathbf{r},\uparrow})$ can be realized by the two Raman beams $\Omega _{1}^{\pi}(\hat{z})=\Omega _{0}e^{ikx}$ and $\Omega _{2}^{x}(\hat{x}+\hat{y})=-\sqrt{2}\Omega _{0}e^{ikz}$ polarized along $(\hat{x}+\hat{y})$-direction, which couple the state $|0\rangle$ and $|2_x\rangle=\left(|\downarrow\rangle+i|\uparrow\rangle\right)/\sqrt{2}$ since $(\hat{x}+\hat{y}) \sim (\sigma^{+}+i\sigma^{-})$. Thus, the hopping term along the $x$ axis $\hat{H}_{\mathbf{rx}}$ can be realized by three Raman beams with the configuration shown in Fig. \ref{RamanBeam} (a). Along the $y$ axis, the hopping term $T_{+y}^{(1)}=- \hat{a}^{\dag}_{\mathbf{r+y},\uparrow}(\hat{a}_{\mathbf{r},\downarrow}-i\hat{a}_{\mathbf{r},0})$ can be realized by three Raman beams $\Omega _{1}^{y}(\sigma^-)=\Omega _{0}e^{iky}$ which excites the state $|\uparrow\rangle$, and $\Omega _{2}^{y}(\sigma^+)=-\sqrt{2}\Omega _{0}e^{ikz}$ and $\Omega _{2}^{\pi}(\hat{z})=i\sqrt{2}\Omega _{0}e^{-iky}$ together excite effectively the state $|1_y\rangle=\left(|\downarrow\rangle-i|0\rangle\right)/\sqrt{2}$. Similarly, the hopping term $T_{-y}^{(2)}=\hat{a}^{\dag}_{\mathbf{r-y},\uparrow}(\hat{a}_{\mathbf{r},\downarrow}+i\hat{a}_{\mathbf{r},0})$ can be realized by two additional Raman beams $\Omega _{3}^{y}(\sigma^+)=\sqrt{2}\Omega _{0}e^{ikz}$ and $\Omega _{3}^{\pi}(\hat{z})=i\sqrt{2}\Omega _{0}e^{-iky}$ which effectively excite the state $|2_y\rangle=\left(|\downarrow\rangle+i|0\rangle\right)/\sqrt{2}$. Note that here a wave-vector difference $\delta \mathbf{k} = (0,-2k)$ and a two-photon energy detuning $\Delta_{y}$ guarantee the hopping along $y$-direction. The laser configuration for realizing the desired hopping term along the $y$ axis $\hat{H}_{\mathbf{ry}}$ is shown in Fig. \ref{RamanBeam} (b). With the eight laser beams required to realize the full Hamiltonian as shown in Fig. \ref{RamanBeam} (c), it is important to forbid the undesired tunneling terms. To this end, we require different linear energy shifts per site $\Delta_x$ and $\Delta_y$ along the $x$ and $y$-directions, which can be achieved by adjusting the direction of the gradient field to be in different angles with respect to the axes of the optical lattice.

\begin{figure}[htbp]\centering
\includegraphics[width=8.5cm]{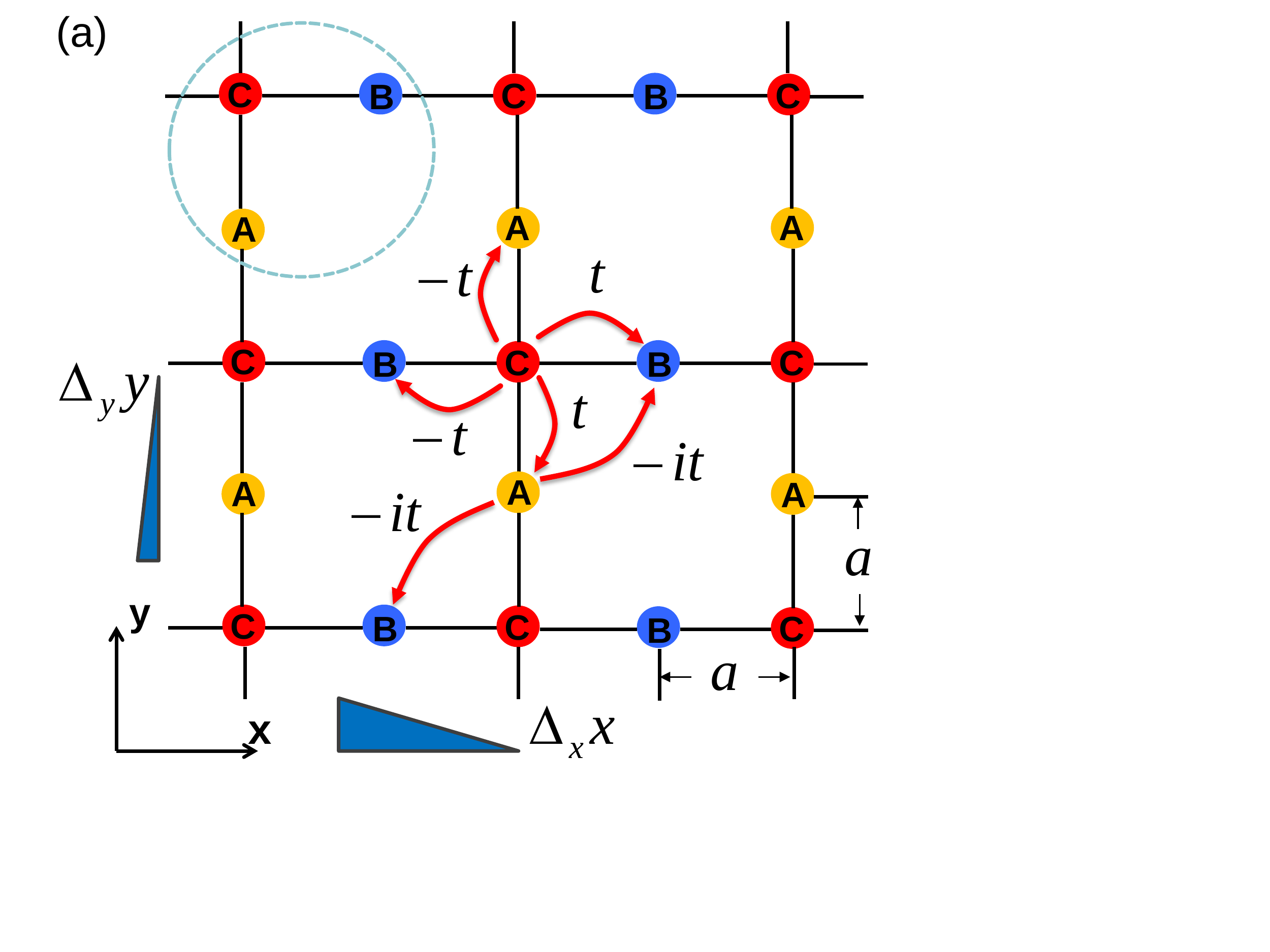}
\includegraphics[width=8.5cm]{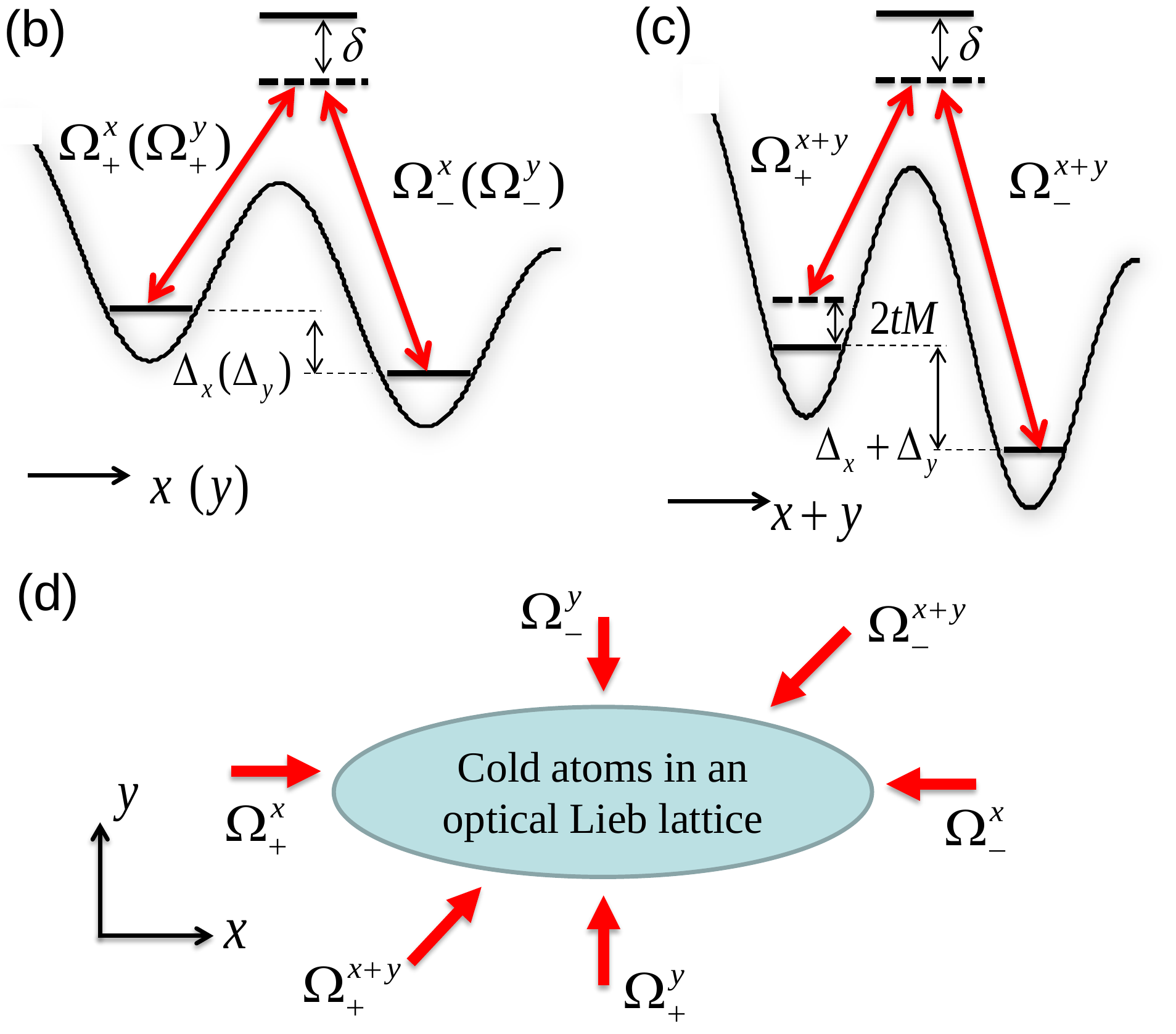}
\caption{(Color online) (a) Schematic diagram of realizing 2D Maxwell points and Maxwell quasiparticles in an optical Lieb lattice. A unit cell indicated by dashed line is composed of three sites labeled by $A$, $B$, $C$, with the lattice constant $a$. The three sublattices form the pseudospin-1 basis, and the spin-flip hopping along each direction with the corresponding hopping amplitude is shown. These hopping can be realized by the Raman-assisted hopping method with the help of the linear title potentials $\Delta_{x}x$ and $\Delta_{y}y$ and the application of laser beams, similar as the scheme in the square optical lattice. (b) Two pairs of Raman beams for inducing the desired hopping along $x$ and $y$ directions; (c) A pair of Raman beams for inducing the desired hopping along $x+y$ direction and a two-photon detuning for inducing the constant term $2tM\hat{S}_z$; (d) The total lasers with the corresponding propagation direction. }\label{LiebLattice}
\end{figure}

In the second part, we point out that the Maxwell points and
the associated Maxwell quasiparticles may be alternatively
realized by using single-component fermionic atoms in optical
lattices with three sublattices, such as an optical Lieb lattice
\cite{Taie}. In experiments, the optical Lieb lattice for cold
atoms has been constructed by superimposing three types of optical
lattices, with the tunable optical potential \cite{Taie}
\begin{equation} \tag{S.15}
\begin{split}
V(x,y)=&-V_{\text{long}}^{x}\cos^2(k_Lx)-V_{\text{long}}^{y}\cos^2(k_Ly)-V_{\text{short}}^{x}\cos^2(2k_Lx)\\
&-V_{\text{short}}^{y}\cos^2(2k_Ly)-V_{\text{diag}}\cos^2\left[2k_L(x-y)+\frac{\pi}{2}\right].
\end{split}
\end{equation}
Here $k_L=2\pi/\lambda$ is a wave number of a long lattice with a depth $V_{\text{long}}$, a short lattice $V_{\text{short}}$ is formed by laser beams at wave length $\lambda/2$, and a diagonal lattice $V_{\text{diag}}$ with the wave number $\sqrt{2}k_L$ is realized by interference of the mutually orthogonal laser beams at $\lambda$ along the $x$ and $y$ directions. The optical Lieb lattice system is shown in Fig. \ref{LiebLattice}(a), with three sublattices $A$, $B$, $C$ forming a unit cell. By tuning the lattice depths $\{V_{\text{long}},V_{\text{short}},V_{\text{diag}}\}$, one can change the energy of the sublattices \cite{Taie}.

In this system, the pseudospin-1 basis are replaced by the three sublattices in a unit cell, and thus the three spin states are given by
\begin{equation} \tag{S.16}
|A\rangle\Longleftrightarrow|\uparrow\rangle, ~|B\rangle\Longleftrightarrow|0\rangle, ~|C\rangle\Longleftrightarrow|\downarrow\rangle.
\end{equation}
In this lattice, the spin-flip hopping $|B\rangle\leftrightarrow|C\rangle$ and $|A\rangle\leftrightarrow|C\rangle$ under the operators $\hat{S}_x$ and $\hat{S}_y$ along the $x$ and $y$ axis become naturally the nearest neighbor hopping in that axis, with the corresponding hopping amplitudes are shown in Fig. \ref{LiebLattice}(a). With the similar Raman-assisted hopping method, the hopping along the $x$ axis and the $y$ axis can be realized by two pairs of laser beams, $\Omega_{\pm}^x=\Omega_0e^{\pm ik_1x}$ and $\Omega_{\pm}^y=\pm\Omega_0e^{\pm ik_1y}$, under the large linear title potentials $\Delta_{x}x$ and $\Delta_{y}y$, respectively, as shown in Fig. \ref{LiebLattice}(b). The detuning in each direction matches the frequency offset of the corresponding Raman beams as we can choose the title energies $\Delta_{x}\approx2.5\Delta_{y}$ with $\Delta_y\gg t_0$ being assumed. In this system, since only one atomic internal state is used in the Raman transitions, then one can address the atoms only through the energy selection without involving the laser polarization \cite{LAT1,LAT2}. Under the two pairs of laser beams, the momenta transferred in the Raman transition along the $x$ and $y$ directions are $\delta \mathbf{k}_1=-2k_1\hat{x}$ and $\delta \mathbf{k}_2=-2k_2\hat{y}$, respectively. Thus the corresponding site-dependent hopping phases along $x$ and $y$ directions are $e^{-2ik_1x}= e^{-2ik_1j_xa}$ and $e^{-2ik_2y}=e^{-2ik_2j_ya}$, with the lattice site index $(j_x,j_y)$. We can choose the parameters $k_1=k_2=\pi/2a$ to induce the hopping phases $e^{-i\pi j_x}=e^{-i\pi j_y}=0,\pi$ staggered along the $x$ and $y$ directions, which lead to the desired hopping $|B\rangle\leftrightarrow|C\rangle$ and $|A\rangle\leftrightarrow|C\rangle$ in the corresponding axis.

The spin-flip hopping $|A\rangle\leftrightarrow|B\rangle$ under
the operator $\hat{S}_z$ in this lattice becomes  next-nearest
neighbor hopping along the $x+y$ or $x-y$ axis, with the
corresponding hopping amplitude along the $x+y$ axis is shown in
Fig. \ref{LiebLattice}(a). This hopping can be achieved by
additional Raman transition by using the third pair of lasers
$\Omega_+^{x+y}=\Omega_0e^{ik_3(x+y)}$ and
$\Omega_-^{x+y}=-i\Omega_0e^{-ik_3(x+y)}$ with a different
matching energy $\Delta_x+\Delta_y=3.5\Delta_y$, as shown in Fig.
\ref{LiebLattice}(c). Here a two-photon detuning in the transition
can be used to induce the constant term $2tM\hat{S}_z$, without
adding other coupling beams in this system. We choose the
parameter $k_3=\pi/a$, then the site dependent phase along the
$x+y$ direction can always be reduced to $e^{-2ik_3(j_x+j_y)a}=1$,
such that the hopping constant $-it$ along this direction is
achieved by the two Raman beams. If the hopping
$|A\rangle\leftrightarrow|B\rangle$ along the $x-y$ axis is
wanted, one can also add the Raman transition with the matching
energy $\Delta_x-\Delta_y=1.5\Delta_y$. The laser configuration of
this system is shown in Fig. \ref{LiebLattice}(d). Under these
conditions, the Bloch Hamiltonian of the 2D Maxwell systems now
becomes
\begin{equation} \tag{S.17}
\begin{split}
\mathcal{H}(\mathbf{k})=&R_x(\mathbf{k})\hat{S}_x+R_y(\mathbf{k})\hat{S}_y+R_z(\mathbf{k})\hat{S}_z,\\
R_x=&2t\sin{k_x},~R_y=2t\sin{k_y},~R_z=2t[M-\cos(k_x+k_y)].
\end{split}
\end{equation}
Here the spin-1 matrices $\hat{S}_{x,y,z}$ acts on the three
sublattices and the lattice constant $a\equiv1$. In this case,
one can obtain the Maxwell points and the associated Maxwell
quasiparticles, similar as the case discussed in the main text.
For instance, when the parameter $M=1$, there is a Maxwell point
at $\mathbf{K}=(0,0)$ with the low-energy effective Hamiltonian
$\mathcal{H}_{\text{eff}}(\mathbf{q})\approx
v\emph{q}_x\hat{S}_x+v\emph{q}_y\hat{S}_y$, where $v=2t$ is the
effective speed of light and $\mathbf{q}=\mathbf{k}-\mathbf{K}$.

\section{Derivation of the topological invariants}

The Berry curvature is given by $\mathbf{F}=\nabla\times\mathbf{A}$, where the Berry connection $\mathbf{A}=-i\langle{\psi}|\mathbf{\nabla}{\psi}\rangle$. For the Bloch Hamiltonian of the 2D model in the main text $\mathcal{H}=\mathbf{R}(\mathbf{k})\cdot\mathbf{S}$, the Berry connection $\mathbf{A}=(A_x,A_y,0)$ for the lowest band with the energy $E=-R$ is given by \cite{He}
\begin{equation}\tag{S.18}
A_\mu=-\frac{R_3}{R(R^2-R_3^2)}(R_2\frac{\partial{R_1}}{\partial{k_\mu}}-R_1\frac{\partial{R_2}}{\partial{k_\mu}}).
\end{equation}
The corresponding Berry curvature is $\mathbf{F}=(0,0,F_{xy})$ with $F_{xy}$ being given by
\begin{equation}\tag{S.19}
\begin{split}
F_{xy}=\frac{\partial{A_y}}{\partial{k_x}}-\frac{\partial{A_x}}{\partial{k_y}}=-\frac{1}{R^3}\varepsilon_{abc}R_{a}\frac{\partial{R_b}}{\partial{k_x}}\frac{\partial{R_c}}{\partial{k_y}}
=-\frac{1}{R^{3}}\mathbf{R}\cdot(\frac{\partial{\mathbf{R}}}{\partial{k_x}}\times\frac{\partial{\mathbf{R}}}{\partial{k_y}}),
\end{split}
\end{equation}
where the Bloch vectors are $R_x=2t\sin{k_x}$, $R_y=2t\sin{k_y}$, and $R_z=2t(M-\cos{k_x}-\cos{k_y})$. A straightforward calculation gives the following form
\begin{equation}\tag{S.20}
F_{xy}=\frac{\cos{k_x}+\cos{k_y}-M\cos k_x\cos k_y}{(\sin^2{k_x}+\sin^2{k_y}+(M-\cos{k_x}-\cos{k_y})^2)^{3/2}}.
\end{equation}
We can thus obtain the Chern number for this band
\begin{equation}\tag{S.21}
\begin{split}
\mathcal{C}=\frac{1}{2\pi}\oint_Sd\mathbf{k}\cdot\mathbf{F(\mathbf{k})}=\frac{1}{2\pi}\oint_S{d^2k}F_{xy}=\left\{\begin{matrix}
2\text{sign}(M), & {0<|M|<2}\\
0, &{|M|>2}
\end{matrix}\right.
\end{split}
\end{equation}

For $M=\pm2$, we respectively expand the Hamiltonian around $\mathbf{K}_+=(0,0)$ and $\mathbf{K}_-=(\pi,\pi)$, and obtain the low-energy effective Hamiltonian
\begin{equation}\tag{S.22}
H_{\pm}(\mathbf{q})=\pm(vq_x\hat{S}_x+vq_y\hat{S}_y-2tm\hat{S}_z),
\end{equation}
where $m=2\mp{M}$, $v=2t$, and $\mathbf{q}=\mathbf{k}-\mathbf{K_\pm}$ with $|\mathbf{q}|\ll|\mathbf{k}|$. We can obtain the effective Berry curvature
\begin{equation}\tag{S.23}
F_{xy}=\pm\frac{m}{(q^2+m^2)^{3/2}},
\end{equation}
where $q=\sqrt{q_x^2+q_y^2}$. Thus the Berry phase $\gamma$ integrated around the Maxwell point $\mathbf{K}_\pm$ for the Fermi surface (FS) can be derived by
\begin{equation}\tag{S.24}
\begin{split}
\gamma=\oint_{FS}d\mathbf{k}\cdot\mathbf{A(\mathbf{k})}=\pm\oint_{FS}{d^2q\frac{m}{(q^2+m^2)^{3/2}}}=\pm\int_0^{2\pi}d\theta\int_{0}^{k_F}\frac{m}{(q^2+m^2)^{3/2}}qdq=\pm{2\pi\int_{0}^{k_F}\frac{m}{(q^2+m^2)^{3/2}}qdq},
\end{split}
\end{equation}
where $k_F$ is the Fermi momentum and the parameter $m\rightarrow0$. Let $q=m\tan{\varphi}$, then we have $1+\tan^2\varphi=\sec^2\varphi$ and $dq=m\sec^2\varphi{d\varphi}$. Substituting these relationships into the above equation, we obtain $\gamma$ as a function of $m$:
\begin{equation}\tag{S.25}
\begin{split}
\gamma=&\pm2\pi\int_{0}^{k_F}\frac{m}{(q^2+m^2)^{3/2}}qdq=\pm2\pi\int_{0}^{\varphi_F}\frac{m^2\tan^2{\varphi}}{m^3\sec^3{\varphi}}m\sec^2{\varphi}d\varphi\\
=&\pm2\pi\int_0^{\varphi_F}\sin{\varphi}d\varphi=\pm2\pi(-\cos\varphi)|_0^{\varphi_F}=\pm2\pi(-\frac{m}{\sqrt{q^2+m^2}})|_0^{k_F}\\
=&\pm2\pi(1-\frac{m}{\sqrt{k^2_F+m^2}}).
\end{split}
\end{equation}
Thus for $m=0$, we obtain $\gamma=\pm2\pi$ for $M=\pm2$.

For $M=0$, we respectively expand the Hamiltonian around $\mathbf{K}_{(0,\pi)}=(0,\pi)$ and $\mathbf{K}_{(\pi,0)}=(\pi,0)$,and obtain the low-energy effective Hamiltonian
\begin{equation}\tag{S.26}
\mathcal{H}_0(\mathbf{q})=\pm({vq_x\hat{S}_x-vq_y\hat{S}_y+2tm_0\hat{S}_z})
\end{equation}
where $m_0=\pm{M}$ in this case, and $\mathbf{q}=\mathbf{k}-\mathbf{K}_{(0,\pi)/(\pi,0)}$, $|\mathbf{q}|\ll|\mathbf{k}|$. We can obtain the Berry connection
\begin{equation}\tag{S.27}
A_x=\pm\frac{m_0q_y}{q^2\sqrt{q^2+m_0^2}}, ~A_y=\pm\frac{m_0q_x}{q^2\sqrt{q^2+m_0^2}}.
\end{equation}
Thus the Berry phase $\gamma$ integrated around the Maxwell point $\mathbf{K}_{(0,\pi)/(\pi,0)}$ for Fermi surface can be derived as
\begin{equation}\tag{S.28}
\begin{split}
\gamma=&\oint_{FS}d\mathbf{k}\cdot\mathbf{A(\mathbf{k})}=\int_0^{2\pi}k_Fd\theta{A_\theta}=\int_0^{2\pi}k_Fd\theta(A_y\frac{q_x}{k_F}-A_x\frac{q_y}{k_F})\\
=&\pm\frac{m_0}{\sqrt{k_F^2+m_0^2}}\int_0^{2\pi}(\cos^2\theta-\sin^2\theta)d\theta=\pm\frac{m_0}{\sqrt{k_F^2+m_0^2}}\int_0^{2\pi}d\theta{\cos(2\theta)}=0,
\end{split}
\end{equation}
where we have used the relationships $q_x=k_F\cos\theta$ and $q_y=k_F\sin\theta$.

\section{A scheme to detect the topological Chern number}

Now we propose a practical method to directly measure the Chern number in our systems, based on a generalization of topological pumping in optical lattices \cite{Thouless,Wang,Zhang1,Marzari,Smith}. We start with the original definition of the Chern number, which is given by
\begin{equation}\tag{S.29}
\begin{split}
\mathcal{C}=\frac{1}{2\pi}\oint{dk}A(\mathbf{k})=\frac{1}{2\pi}\int^{\pi}_{-\pi}dk_{y}\partial_{k_y}\int^{\pi}_{-\pi}{dk_x}A(k_x,k_y)=\frac{1}{2\pi}\int^{\pi}_{-\pi}dk_{y}\partial_{k_y}\varphi_{Zak}(k_y)\\
\end{split}
\end{equation}
where $A(k)=-i\langle{u(k)}|\partial_k|u(k)\rangle$ is the Berry connection, and the Zak phase for a reduced 1D system is $\varphi_{Zak}=\int^{\pi}_{-\pi}{dk}A(k)=2\pi\langle{w}|\hat{x}|{w}\rangle=2\pi\langle{n_x}\rangle$. Here $|w\rangle$ denotes the Wannier functions. We can obtain the Chern number of our 2D Maxwell insulators
\begin{equation}\tag{S.30}
\mathcal{C}=\int^{\pi}_{-\pi}dk\partial_{k_y}\langle{n_x(k_y)}\rangle
\end{equation}
$\langle{n_x(k_y)}\rangle$ is the center of hybrid Wannier
function (HWF) \cite{Wang,Zhang1,Marzari,Smith}.  The HWF center
for tight-binding chain in our system with lattice site $L_x$
under open boundary condition is given by
\begin{equation}\tag{S.31}
\langle{n_x(k_y)}\rangle=\frac{\sum_{i_x}i_{x}\rho(i_x,k_y)}{\sum_{i_x}\rho(i_x,k_y)}
\end{equation}
where $i_x$ is the lattice-site index, and $\rho(i_x,k_y)$ is the
density of the HWF and denotes the atomic densities  resolved
along $x$ direction as a function of $k_y$. Here, the hybrid
density can be written as
\begin{equation}\tag{S.32}
\rho(i_x,k_y)=\sum_{occupied\
{states}}|{i_x,k_y}\rangle\langle{i_x,k_y}|,
\end{equation}
where $|{i_x,k_y}\rangle$ is the hybrid eigenstates of the system. We can prove that the Chern number can be obtained from the shift of the HWF:
\begin{equation}\tag{S.33}
\mathcal{C}=\sum_{j}\Delta\langle{n_{x}(k_y(j))\rangle}
\end{equation}
where $\Delta\langle{n_{x}(k_y(j))}\rangle$ represent the
difference of the HWF center $\langle{n_x}\rangle$ at  the jump
discontinuity $k_y(j)$, $j$ is the number of jump discontinuity
for $\langle{n_x}\rangle$, the sum of the difference at all jump
points is equal to the Chern number.\par
$\mathbf{Prove}$: Due to the system is under period boundary condition, $\langle{n_x(\pi)}\rangle=\langle{n_x(-\pi)}\rangle$.\\
$1$. While $j=0$, $\langle{n_x}\rangle$ is continuous.
\begin{equation}\nonumber
\mathcal{C}=\int^{\pi}_{-\pi}dk_{y}\partial_{k_y}\langle{n_x(k_y)}\rangle=\int^{\pi}_{-\pi}d\langle{n_x(k_y)}\rangle=\langle{n_x(k_y)}\rangle|^{\pi}_{-\pi}=0
\end{equation}\\
$2$. While $j\neq0$, $\langle{n_x}\rangle$ is discontinuous. For
simply, we suppose function $f(x)$  have $j$ times jump
discontinuities, divided $f(x)$ into $j+1$ segment, and satisfied
$f(x_0)=f(x_{j+1})$, $x(j)$ is the jump discontinuity. Then we
have
\begin{equation}\nonumber
\begin{split}
\int^{x_{j+1}}_{x_0}dx\partial_{x}f(x)&=\int^{x_{j+1}}_{x_0}df(x)=\int^{x_{1}}_{x_0}df(x)+\int^{x_{2}}_{x_1}df(x)+\int^{x_{3}}_{x_2}df(x)+\cdots+\int^{x_{j}}_{x_{j-1}}df(x)+\int^{x_{j+1}}_{x_j}df(x)\\
&=f(x)|^{x_1}_{x_0}+f(x)|^{x_2}_{x_1}+f(x)|^{x_3}_{x_2}+\cdots+f(x)|^{x_j}_{x_{j-1}}+f(x)|^{x_{j+1}}_{x_j}\\
&=f(x_1)'-f(x_0)+f(x_2)'-f(x_1)+f(x_3)'-f(x_2)+\cdots+f(x_j)'-f(x_{j-1})+f(x_{j+1})-f(x_j)\\
&=f(x_{j+1})-f(x_0)+f(x_1)'-f(x_1)+\cdots+f(x_j)'-f(x_j)=\sum_{j}(f(x_j)'-f(x_j))
\end{split}
\end{equation}
where $f(x_j)'$ and $f(x_j)$ are belong to the $j$th and $(j+1)$th segment of $f(x)$ at $j$th jump discontinuity, respectively.
\begin{figure}[htbp]\centering
\includegraphics[width=10cm]{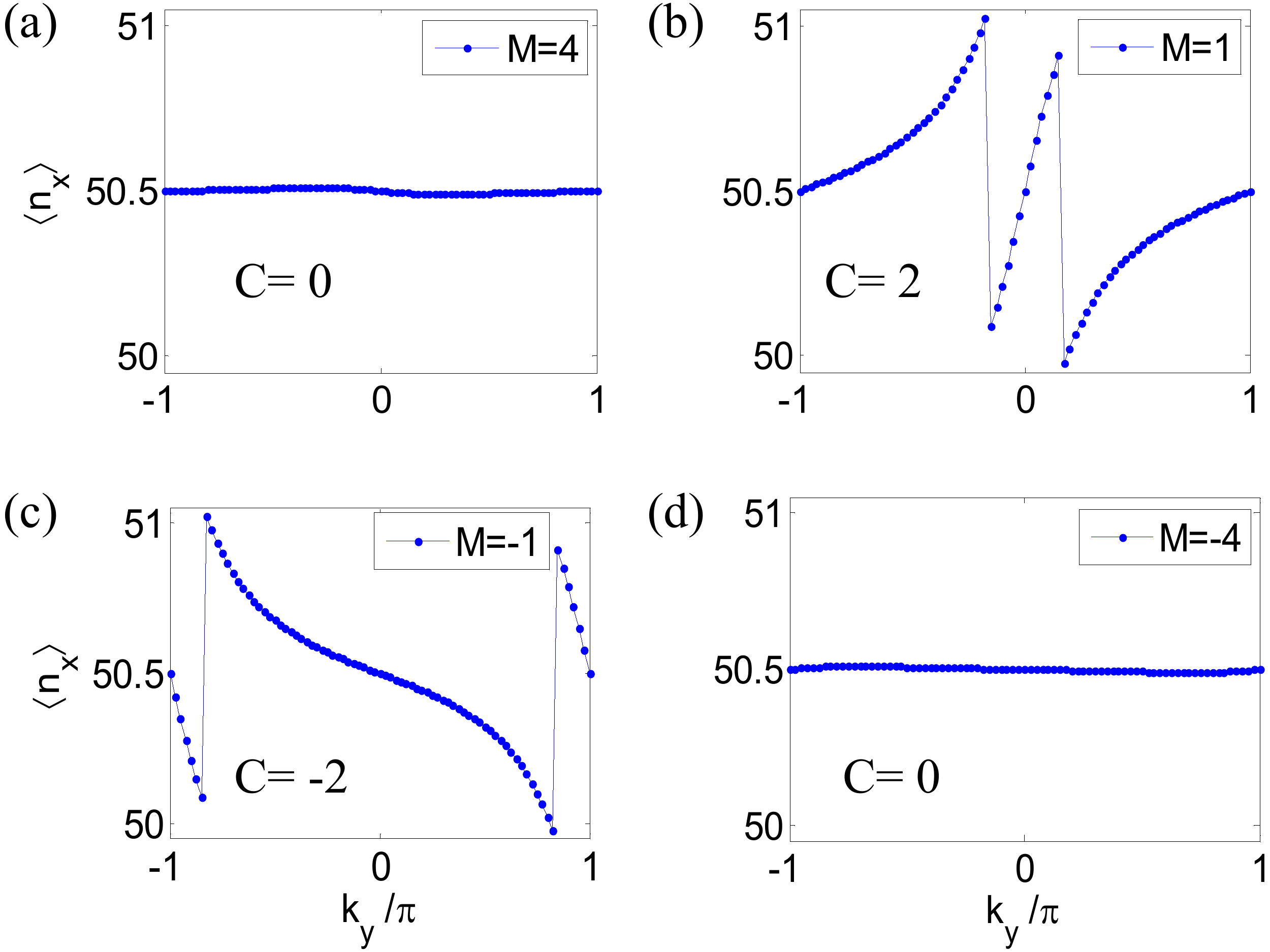}
\caption{(Color online) The HWF centers in a tight-binding chain of length $L_x=100$ at $1/3$ filling as a function of the adiabatic pumping parameter $k_y$ for different parameters $M$. There is no jump of $\langle{n_{x}(k_y)}\rangle$ in (a) and (d), which is consistent with the expected $\mathcal{C}=0$ for $M>2$ and $M<-2$ in these cases. In (b) and (c), $\langle{n_{x}(k_y)}\rangle$ both show the jump of two unit cell for $M=1$ and $M=-1$, corresponding to nontrivial cases with $\mathcal{C}=2$ and $\mathcal{C}=-2$, respectively. } \label{2DHWfs}
\end{figure}

According to these results, we describe a scheme to directly detect the Chern number based on a generalization of topological pumping in optical lattices \cite{Thouless,Wang,Zhang1,Marzari,Smith}. Our 2D insulating Hamiltonian can be viewed as a fictitious 1D insulator subject to an external parameter $k_y$, and we also know the polarization of this 1D insulator can be expressed by means of the center of the HWFs \cite{Wang,Zhang1,Marzari,Smith}, which are localized in the $x$ axis retaining Bloch character in the $k_y$ dimension in our case. This polarization is a function of $k_y$, which acts as an external parameter under which the polarization changes. When $k_y$ is adiabatically changed by $2\pi$, the change in polarization, i.e., the shift of the HWF center, is proportional to the Chern number. This is a manifestation of topological pumping \cite{Thouless}, with $k_y$ being the adiabatic pumping parameter. In Fig. \ref{2DHWfs}, we numerically calculate $\langle{n_{x}(k_y)}\rangle$ in a tight-binding chain of length $L_x=100$ at $1/3$ filling (assuming the Fermi energy $E_F=0$). For the case in Figs. \ref{2DHWfs}(a) and \ref{2DHWfs}(d), $\langle{n_{x}(k_y)}\rangle$ shows no jump, which are consistent with the expected $\mathcal{C}=0$ for the trivial case $M=4$ and $M=-4$. The results for $M=1$ in Fig. \ref{2DHWfs}(b) shows two discontinuous jumps of one unit cell, indicating that a particle is pumped across the system \cite{Marzari,Smith}, the Chern number for this case is $\mathcal{C}=2$, and the result in Fig. \ref{2DHWfs}(c) is similar to Fig. \ref{2DHWfs}(b) but with opposite jump direction and Chern number for $M=-1$. This establishes a direct and clear connection between the shift of the hybrid density center and the topological invariant.


In cold atom experiments, the atomic density $\rho(i_x,k_y)$ can
be directly measured by the hybrid time-of-flight images
\cite{Wang},  which is referring to an \textit{in situ}
measurement of the density distribution of the atomic cloud in the
$x$ direction during free expansion along the $y$ direction. In
the measurement, the optical lattice is switched off along the $y$
direction while keeping the system unchanged in the $x$ direction.
One can map out the crystal momentum distribution along $k_y$ in
the time-of-flight images and a real space density resolution in
the $x$ direction can be done at the same time. Thus one can
directly extract Chern number from this hybrid time-of-fight
images in the cold atom system.

\end{document}